\documentclass[conference]{IEEEtran}
\IEEEoverridecommandlockouts
\usepackage{cite}
\usepackage{amsmath,amssymb,amsfonts}
\usepackage{algorithmic}
\usepackage{graphicx}
\usepackage{textcomp}
\usepackage{xcolor}
\usepackage{subcaption}
\usepackage{hyperref}
\usepackage{rotating}
\def\BibTeX{{\rm B\kern-.05em{\sc i\kern-.025em b}\kern-.08em
    T\kern-.1667em\lower.7ex\hbox{E}\kern-.125emX}}
\begin{document}

\title{Analyzing X's Web of Influence: Dissecting News Sharing Dynamics through Credibility and Popularity with Transfer Entropy and Multiplex Network Measures\\ 
\thanks{This work was partially supported by the Defense Advanced Research Projects Agency (DARPA) under agreement HR00112290104 (PA-21-04-06)}
}

\author{\IEEEauthorblockN{1\textsuperscript{st} Sina Abdidizaji}
\IEEEauthorblockA{\textit{Department of Industrial Engineering} \\
\textit{University of Central Florida}\\
Orlando, FL, USA \\
sina.abdidizaji@ucf.edu}
\and
\IEEEauthorblockN{2\textsuperscript{nd} Alexander Baekey}
\IEEEauthorblockA{\textit{Department of Computer Science} \\
\textit{University of Central Florida}\\
Orlando, FL, USA \\
alexander.baekey@ucf.edu}
\and
\IEEEauthorblockN{3\textsuperscript{rd} Chathura Jayalath}
\IEEEauthorblockA{\textit{Department of Industrial Engineering} \\
\textit{University of Central Florida}\\
Orlando, FL, USA \\
chathura@ucf.edu}
\and
\IEEEauthorblockN{4\textsuperscript{th} Alexander Mantzaris}
\IEEEauthorblockA{\textit{Department of Statistics and Data Science} \\
\textit{University of Central Florida}\\
Orlando, FL, USA \\
alexander.mantzaris@ucf.edu}
\and
\IEEEauthorblockN{5\textsuperscript{th} Ozlem Ozmen Garibay}
\IEEEauthorblockA{\textit{Department of Industrial Engineering} \\
\textit{University of Central Florida}\\
Orlando, FL, USA \\
ozlem@ucf.edu}
\and
\IEEEauthorblockN{6\textsuperscript{th} Ivan Garibay}
\IEEEauthorblockA{\textit{Department of Industrial Engineering} \\
\textit{University of Central Florida}\\
Orlando, FL, USA \\
igaribay@ucf.edu}
}

\maketitle

\begin{abstract}
The dissemination of news articles on social media platforms significantly impacts the public's perception of global issues, with the nature of these articles varying in credibility and popularity. The challenge of measuring this influence and identifying key propagators is formidable. Traditional graph-based metrics such as different centrality measures and node degree methods offer some insights into information flow but prove insufficient for identifying hidden influencers in large-scale social media networks such as X (previously known as Twitter). This study adopts and enhances a non-parametric framework based on Transfer Entropy to elucidate the influence relationships among X users. It further categorizes the distribution of influence exerted by these actors through the innovative use of multiplex network measures within a social media context, aiming to pinpoint influential actors during significant world events. The methodology was applied to three distinct events, and the findings revealed that actors in different events leveraged different types of news articles and influenced distinct sets of actors based on the news category. Notably, we found that actors disseminating trustworthy news articles to influence others occasionally resort to untrustworthy sources. However, the converse scenario, wherein actors predominantly using untrustworthy news types switch to trustworthy sources for influence, is less prevalent. This asymmetry suggests a discernible pattern in the strategic use of news articles for influence across social media networks, highlighting the nuanced roles of trustworthiness and popularity in the spread of information and influence. 
\end{abstract}


\section{Introduction}
In the contemporary era, individuals frequently opt to consume the latest news in digital format. Given the constraints of time and the impracticality of checking numerous news outlets, many turn to social media platforms to stay informed about the latest developments within their areas of interest. Recognizing this trend, manipulative actors can exploit social media to propagate selected news articles, thus influencing the perceptions and behaviors of others, leading them to question the reliability of their traditional news sources and potentially alter their preferences. Bovet and Makse \cite{bovetInfluenceFakeNews2019} analyzed approximately 30 million tweets from 2.2 million users on X (formerly Twitter) over a five-month period preceding the 2016 US presidential election, discovering that about 25\% of these tweets constituted fake news. Another study \cite{madrid-moralesWhoSetNarrative2021} investigated the influence and presence of Chinese news regarding COVID-19 in African countries' news outlets, revealing that former colonial powers exerted more influence over the media content of African countries than Chinese media.

The concept of an influencer within social media varies across different research domains. Generally, an influencer is defined as an individual capable of affecting the opinions and behaviors of others \cite{rodriguez-vidalAutomaticDetectionInfluencers2019a}. In the context of social media, influencers are often identified by their extensive networks, quantified through graph structures and social networks as nodes(users) with numerous edges(connections) \cite{jainOpinionLeaderDetection2020,jainDiscoverOpinionLeader2019}. Huynh et al.\cite{huynhMeasuresDetectInfluencer2019}, however, define influencers based on the volume of posts they publish or republish. Contrary to the assumption that a high number of interactions and followers automatically signifies substantial influence, evidence suggests that these metrics do not always accurately measure a user's impact within social media \cite{deveirmanMarketingInstagramInfluencers2017}. In certain cases, individuals with smaller followings may exert more substantial influence \cite{zareiCharacterisingDetectingSponsored2020a}. Influence within social networks can propagate like a chain, exemplified by sequences of retweets \cite{bhowmickTemporalSequenceRetweets2019}. An influence cascade refers to the sequence of user actions within a social network, beginning with an initial user acting due to external stimuli and leading to subsequent users being influenced to act, continuing until the chain of influence ceases. Essentially, it encompasses all users and events within a social network that are traced back to an initial action initiated by external motivation rather than social influence \cite{senevirathnaInfluenceCascadesEntropyBased2021}. In this research, an influential actor is someone using a specific type of news articles and influencing others to change their taste and start using the same type of the article to propagate the news and creating a cascade of influence until the chain of influence ends. This inquiry is predicated on the hypothesis that the patterns of influence among actors in a social network are significantly shaped by the credibility and popularity of the news sources they elect to share. It is important to clarify that our data collection does not include metrics such as tweet counts, retweets, or followers, yet analyzing tweet frequency and content has enabled us to identify influencers.

Influence campaigns, critical to government agencies due to their global impact, challenge traditional social network analysis methods such as graph metrics, which only partly capture influencer dynamics \cite{kitsakIdentificationInfluentialSpreaders2010}. This research utilizes transfer entropy networks \cite{schreiberMeasuringInformationTransfer2000} to reveal hidden influencers, those with significant but not obvious influence, by focusing on the quantitative impact of shared content on user opinions rather than simple information flow. Expanding on previous studies limited to cross-platform influence \cite{wangMultiPlatformAnalysisPolitical2021}, this paper uses multiplex network measures to characterize influential actors and delineate their relationships within a Transfer Entropy Network on X, and tests them on three major events: the Skripal assassination, the Ukraine war, and the Navalny death.

Following the identification of influential actors and their targets, this study has several contributions aimed at elucidating the dynamics of influence within social networks. Specifically, the research seeks to:

1. Analyze the distribution of influence exerted by actors categorized based on their engagement with news sources of varying credibility and popularity. These categories include those who disseminate content from Trustworthy Mainstream (TM), Trustworthy Fringe (TF), Untrustworthy Mainstream (UM), and Untrustworthy Fringe (UF) news sources. The aim is to understand how influence is distributed by actors using these different source types among targets

2. Compare the behavior of actors in terms of propagating news articles and influencing their targets with those news articles across two different assassination events, and contrast these with a war event to observe the differences in the types of news articles used to disseminate information on X

3. Investigate the extent to which actors engage in disseminating content from more than one type of influence source, thereby acting as conduits for multiple streams of information

The structure of this paper is organized as follows: Section 2 provides a comprehensive review of the literature pertinent to our research. Subsequently, Section 3 details the data collection process and the methodology employed. In Section 4, the results are discussed extensively. Finally, in the concluding section, we summarize the findings and underscore the main contributions of this study.

\section{Previous work}
The phenomenon of influence spread through cascades within social networks has garnered substantial interest across various disciplines, including marketing, sociology, and political science. Analyzing influence in a social network is a complex problem with potential for multiple interpretations and no ground truth. Seyfosadat et al. \cite{seyfosadatSystematicLiteratureReview2023b} identify four primary methodologies for identifying influence within social networks: data mining , machine learning, meta-heuristic approaches, graph-based methods, and a hybrid approach that combines these techniques. Graph-based methods, which is favored in this research domain \cite{seyfosadatSystematicLiteratureReview2023b}, focus on the structural metrics of a graph combined with mathematical models to identify the key influential entities within a social network. While graph theory and network analysis techniques are commonly employed to extract valuable insights about the nodes and edges in a social network and assess influence \cite{pengInfluenceAnalysisSocial2018a}, platform-specific metrics such as retweets and follower counts also serve as criteria for evaluating influence \cite{singhIdentificationInfluencePropagation2019a}. However, these straightforward metrics offer limited scope; for instance, a high in-degree does not necessarily equate to significant influence \cite{chaMeasuringUserInfluence2010a}. Furthermore, Kitsak et al. \cite{kitsakIdentificationInfluentialSpreaders2010} discovered that the most pivotal influencers are not always those with the most connections but are rather strategically positioned at the core of the network structure. This underscores the necessity for advanced techniques that transcend mere connection counts to accurately identify influencers.

Entropy-based approaches represent a promising avenue for assessing influence within social networks. 
Saxena et al. \cite{saxenaEntropyBasedFlow2020a} developed the Entropy-based Influence Disseminator (EbID) method, which uses an entropy-based centrality measure to identify key spreaders in networks by evaluating the entropy of influence paths and community attributes of neighboring nodes for wider spread potential. Their model outperformed others relying solely on centrality measures. He et al. \cite{heIdentifyingPeerInfluence2013} developed a model-free methodology to derive causal inferences about user behavior in social networks, employing Transfer Entropy to uncover both implicit and explicit causal relationships. Steeg and Galstyan \cite{versteegInformationtheoreticMeasuresInfluence2013} proposed the concept of content transfer, a metric grounded in information theory that offers predictive insights by quantifying the impact of one user's content on another, without dependency on predefined models. This approach uses non-parametric entropy calculations and sophisticated content representation techniques to identify predictive relationships among users, even in the absence of direct connections through following or mentions. Subsequently, Senevirathna et al. \cite{senevirathnaInfluenceCascadesEntropyBased2021} further advanced the transfer entropy model by introducing the influence cascade model, wherein a node initiates influence that impacts another node, which then propagates this influence further within the social network. This model allows for the tracing of influence back to its origin node. Additionally, they devised a method to visualize the direction of entropy flows between nodes, termed an influence cascade. This framework was further expanded into the Influence Cascades Ecosystem \cite{garibayEntropyBasedCharacterizationInfluence2022a}, applying it to a geopolitical news model to trace influence across both traditional and social networks within a hybrid environment.

Within the realm of network science, Multiplex Networks have emerged as a powerful tool for tackling a wide array of network-centric challenges. Often, analyzing a solitary network falls short of providing comprehensive insights, whereas the construction of a multiplex network — comprising connected single networks as layers — facilitates the extraction of valuable information \cite{battistonStructuralMeasuresMultiplex2014a}. Their analytical use is heavily dependent on the data environment and the type of connection between nodes in the network, creating distinct layers of the multiplex networks. They have been used recently for modeling complex network systems such as information diffusion  \cite{liHowMultipleSocial2015a}, disease characterization \cite{haluMultiplexNetworkHuman2019a} and so forth. Furthermore, they have been applied to evaluate social ties \cite{hristovaKeepYourFriends2014a} and facilitate knowledge dissemination \cite{zhuResearchKnowledgeDissemination2022a} within social networks. This study proposes a novel application of multiplex networks for characterizing influence within social networks, specifically focusing on X. This method can be advantageous when the analysis of a single network of influence fails to yield meaningful information, and the aggregation with other networks provides insights into the interactions between key influential players across different networks. We build upon the foundational work of Battiston et al. \cite{battistonStructuralMeasuresMultiplex2014a} and integrate transfer entropy to enhance our understanding of node behavior and influence within a multiplexed network setting.

\section{Methods}
\begin{figure*}[htbp]
    \centering
    \includegraphics[width=140mm]{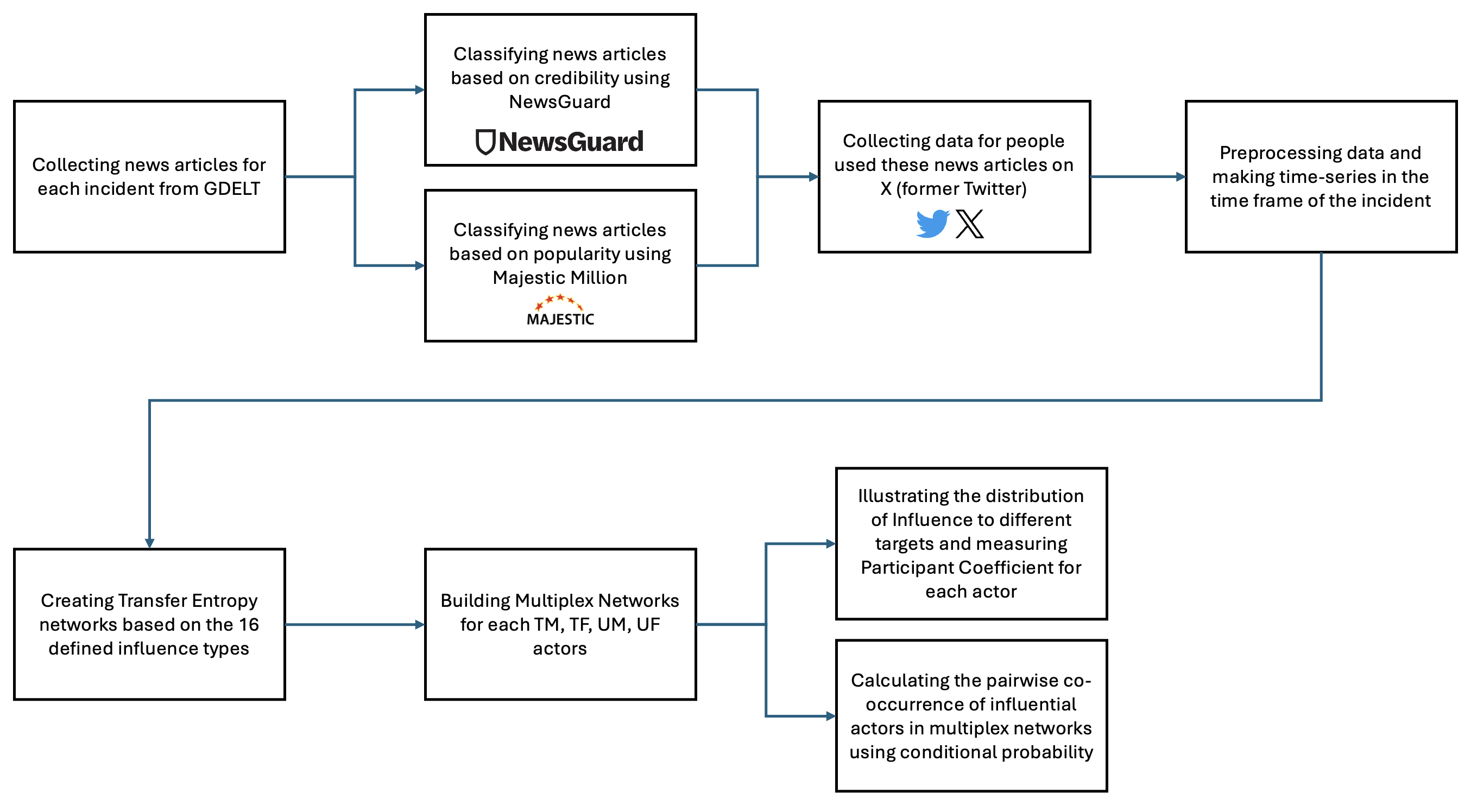}
    \caption{The full framework pipeline of transfer entropy with multiplex networks for analyzing influential actors' behavior}
    \label{fig1}
\end{figure*}
\subsection{Data collection and Preparation}
For data collection, this study selected three events for investigation. Data pertaining to these events was systematically gathered using the Brandwatch platform \footnote{https://www.brandwatch.com/}. The scope of data collection focused on profiles on X that shared news articles related to the chosen events. The first event pertains to the assassination attempt and subsequent poisoning of Sergei Skripal, with data for this incident being collected from March 1, 2018, to May 4, 2018. The second event involves the conflict between Ukraine and Russia, with the corresponding data collection spanning from January 1, 2022, to May 1, 2022. The third event concerns the suspicious death of Alexei Navalny, for which data was gathered from February 12, 2024, to March 2, 2024. Subsequently, these news articles were subjected to a classification process based on their credibility and popularity. Extending the framework introduced by Wang \cite{wangMultiPlatformAnalysisPolitical2021}, we classify news articles by trustworthiness and popularity based on NewsGuard \footnote{https://www.newsguardtech.com} and Majestic Million \footnote{https://majestic.com/reports/majestic-million} rankings, respectively. That said, these rankings can be replaced by any desired classification scheme. To evaluate the trustworthiness of news articles, NewsGuard assigns scores to news agencies indicative of their credibility. Articles from sources scoring above 60 were designated as trustworthy, whereas those with scores below 60 were classified as untrustworthy. The classification of news articles by popularity conducted by Majestic Million database allowed for the categorization of articles into Mainstream, representing those from highly popular websites, and Fringe, denoting articles from less popular sources. Such categorization is critical for understanding the diversity of information sources and facilitates a nuanced analysis of information dissemination practices on social media. After conducting the classification, the time series data of actors in X was preprocessed to calculate the Transfer Entropy networks.
\subsection{Transfer Entropy}
Transfer Entropy is a statistical measure of information transfer introduced by Schreiber \cite{schreiberMeasuringInformationTransfer2000} which is an information-theoretical measure derived from Shannon entropy. TE is used to evaluate the extent to which knowing the historical data of two random processes (denoted as $X_t$ and $Y_t$) can reduce the uncertainty in predicting the future state of one process, effectively quantifying the influence or informational flow from one process to the other. The TE metric is directional and non-commutative, indicating that the transfer of information from $X$ to $Y$ can differ in magnitude from $Y$ to $X$.

In the mathematical formulation, as given in equation\ref{eq:te}, TE from $X$ to $Y$ ($\text{TE}_{X\rightarrow Y}$) is calculated using a specific equation that involves summing over the probabilities of the next state of $Y$ ($Y_{t+1}$), given its own history and the history of 
$X$, and then comparing this probability to the probability of $Y_{t+1}$ given its own history alone. In this study the histories of $X$ and $Y$ are denoted by $Y_t$ and $X_t$ which represent sequences of past observations of length one. 

\begin{equation}
\label{eq:te}
    TE_{X \rightarrow Y} = \sum P(Y_{t+1},Y_{t},X_{t}) \log\frac{P(Y_{t+1}|Y_{t},X_{t})}{P(Y_{t+1}|Y_{t})} 
\end{equation}

The application of TE is demonstrated in a study involving user data on X. For each user, the study extracts a time series of their tweeting activity, identifying all timestamps of their tweets. The data is then resampled at a daily frequency ($f=1$ Day), converting the activity into a binary time series where a "1" represents a day the user tweeted, and a "0" signifies a day they did not. This binary time series is then applied to the TE calculation to analyze the information flow between users based on their tweeting activity.

The weight assigned to an edge within a TE network is indicative of the strength of influence exerted within this network. Four specific types of influence were delineated, based on the actors' connections to Trustworthy Mainstream (TM), Untrustworthy Mainstream (UM), Trustworthy Fringe (TF), and Untrustworthy Fringe (UF) news sources. For instance, a TM actor might target four different audience groups: those engaging with TM news sources, indicative of echo chambers; TF targets, representing popularity crossover; UM targets, signifying trust crossover; and UF targets, illustrating both trust and popularity crossovers. The full description is provided in \autoref{tab1}.

\begin{table}[htbp]
\centering
\caption{Definition of 16 influence types based on credibility and popularity of news articles}
\label{tab1}
\begin{tabular}{ll}
\hline
{\textbf{Name}}& \multicolumn{1}{l}{{\textbf{Class}}} \\ \hline
{TM$\rightarrow$TM} & {echochamber} \\ 
{TF$\rightarrow$TF} & {echochamber} \\ 
{UM$\rightarrow$UM} & {echochamber} \\ 
{UF$\rightarrow$UF} & {echochamber} \\ 
{TM$\rightarrow$UM} & {credibility crossover} \\ 
{TF$\rightarrow$UF} & {credibility crossover} \\ 
{UM$\rightarrow$TM} & {credibility crossover} \\ 
{UF$\rightarrow$TF} & {credibility crossover} \\ 
{TM$\rightarrow$TF} & {audience crossover} \\ 
{UM$\rightarrow$UF} & {audience crossover} \\ 
{TF$\rightarrow$TM} & {audience crossover} \\ 
{UF$\rightarrow$UM} & {audience crossover} \\ 
{TM$\rightarrow$UF} & {credibility and audience crossover} \\ 
{TF$\rightarrow$UM} & {credibility and audience crossover} \\ 
{UM$\rightarrow$TF} & {credibility and audience crossover} \\ 
{UF$\rightarrow$TM} & {credibility and audience crossover} \\ \hline
\end{tabular}
\end{table}

\subsection{Multiplex Networks and Aggregates}
\begin{figure}[htbp]
    \centering
    \includegraphics[width= 8cm, height=8cm]{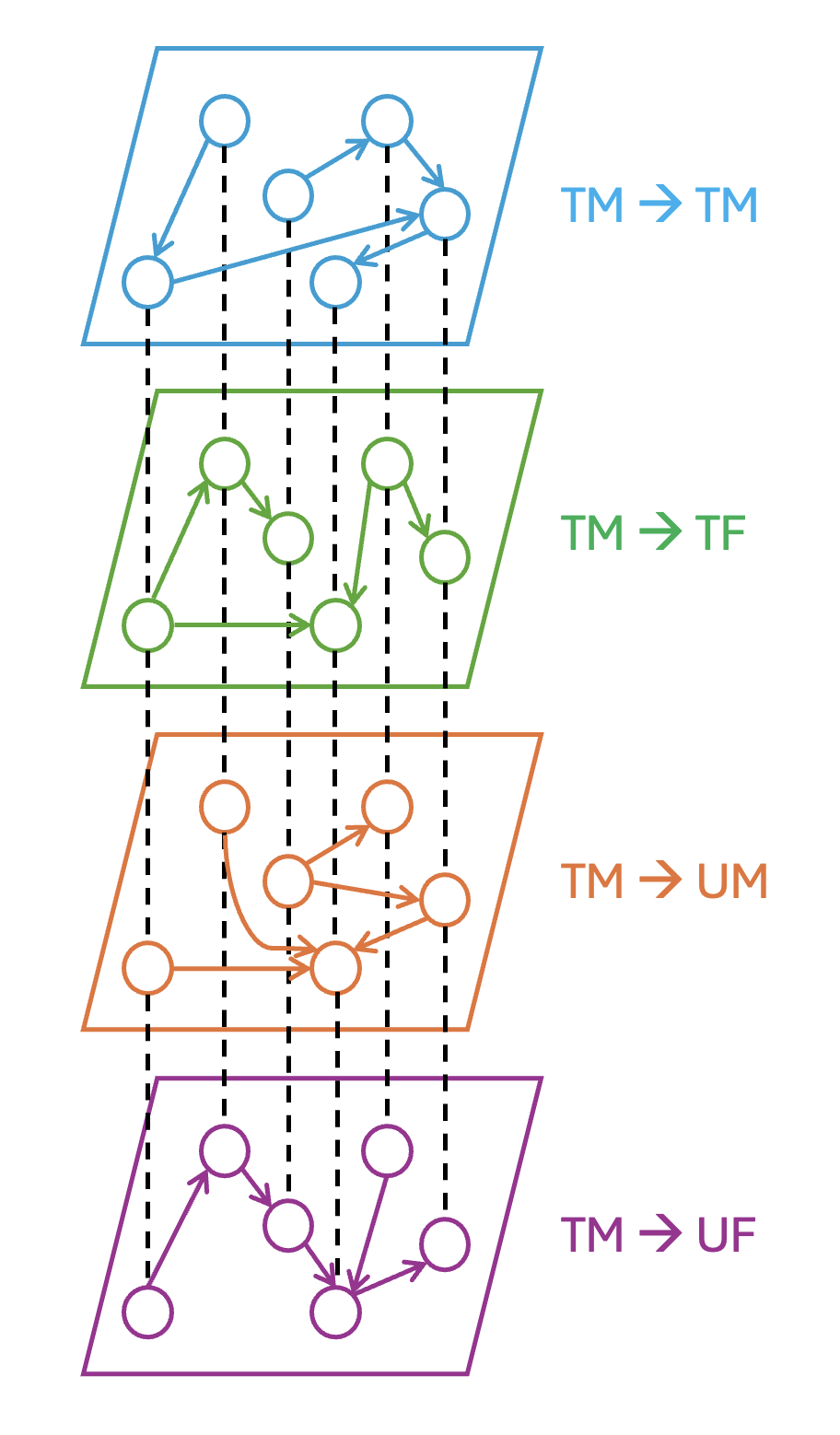}
    \caption{Construction of the Multiplex Network for TM Actors Across Four Layers Demonstrating the Direction of Influence Towards Various Types of Targets}
    \label{fig2}
\end{figure}

Taking the necessary step of building a TE network opens up many analytical possibilities. The most obvious of which being a variety of aggregations by constructing TE networks with the defined classification. Aggregating edge types offers a number of perspectives that are inaccessible until the TE network is built. Upon the construction of Transfer Entropy, a critical endeavor entailed the analysis of influential actors distribution across varied types of influence within these networks. To achieve this purpose, we need to build multiplex networks for different types of influential actors. For instance, from the 16 edge types in the TE graph, one can aggregate over all ${TM\rightarrow **}$ edges (${TM\rightarrow TM}$, ${TM\rightarrow TF}$, ${TM\rightarrow UM}$, ${TM\rightarrow UF}$) which will show all of the influence from sources that share Trustworthy Mainstream news articles and affects people shared other type of news article, causing to form a multiplex network by combining links between actors in these 4 layers. The creation of this TM multiplex network is depicted in \autoref{fig2}.
In mathematical terms, to construct the multiplex network, the aggregate transfer entropy for each node within a single TE network is computed as follows \cite{battistonStructuralMeasuresMultiplex2014a}:
\begin{equation}
k_i = \sum_j a_{ij}
\end{equation}
Here, \(k_{i}\) represents the total transfer entropy emitted from node \(i\) within the TE network, where \(a_{ij}\) denotes the directed edge from node \(i\) to node \(j\), indicating the influence exerted by node \(i\) on node \(j\). This analysis exclusively considers outgoing directed edges, as the focus is on the exertion of influence rather than the reception thereof. Following the determination of total transfer entropy within individual TE networks, multiplex networks for TM, UM, TF, and UF actors are constructed. This is mathematically expressed as \cite{battistonStructuralMeasuresMultiplex2014a}:
\begin{equation}
o_i = \sum_\alpha k_{i}^{[\alpha]}
\end{equation}
The equation above calculates the total transfer entropy for a node across the multiplex network, summing the outgoing edges of that node across all single TE networks denoted by \(\alpha\). 

\subsection{Participation Coefficient}
The approach of creating multiplex networks facilitated an in-depth examination of how influence distribution among actors permeates through various layers within the multiplex networks. To quantitatively assess the extent of influence dispersion among actors, multiplex network measures were applied, notably the Participant Coefficient \cite{battistonStructuralMeasuresMultiplex2014a}. The calculation of this coefficient uses the formula below:
\begin{equation}
P_{i}=\frac{M}{M-1}[1-\sum_{\alpha=1}^{M}(\frac{k_{i}^{[\alpha]}}{o_{i}})^{2}]
\end{equation}
This metric quantitatively assesses the extent of actor involvement across distinct layers within multiplex networks. Within the context of this formulation, \(M\) signifies the total number of layers present in each multiplex network. The variable \(\alpha\) represents a layer within the TE networks. The term \(k_{i}^{[\alpha]}\) denotes the aggregated transfer entropy associated with a node within a singular TE network layer, while \(o_{i}\) encapsulates the cumulative transfer entropy attributed to a node across the entirety of the multiplex network. A low participant coefficient, approaching zero, signifies that an actor's influence is predominantly directed at a singular audience type. Conversely, a high participant coefficient, nearing one, indicates an equal distribution of influence across multiple audience layers.

\subsection{Conditional Probability for pairwise co-occurrence}
In the next phase, the goal was to determine the percentage of actors active in different influence types simultaneously. As observed, some actors were identified in different influence pathways, prompting an investigation into whether these actors employ different news links in terms of trustworthiness and popularity in their tweets or if they propagate one type of influence through diverse news sources. The aim of this evaluation was to see if actors targeted only one type of influence to propagate in the network or if they were active in multiple types of influence. These established multiplex networks underwent pairwise comparisons. To calculate it, the formula below was used \cite{battistonStructuralMeasuresMultiplex2014a}:
\begin{equation}
P\left(a^{[\alpha']}_{ij} \middle| a^{[\alpha]}_{ij}\right) = \frac{\sum_{ij} a^{[\alpha']}_{ij} a^{[\alpha]}_{ij}}{\sum_{ij} a^{[\alpha]}_{ij}}
\end{equation}
This formula involved assessing whether an actor, active and categorized in one type of influence, was also active in another influence type or not. These comparisons provided valuable insights into the cross-influence patterns among actors in our study.

To the best of our knowledge, this study represents the inaugural application of this particular classification combined with Transfer Entropy to examine influential actors on a social media platform \footnote{Code: \url{https://github.com/sina6990/Multiplex-Networks}}. Furthermore, the employment of multiplex networks with such a distinction has not been documented in previous research focusing on the study of influential actors. The full framework pipeline is illustrated in \autoref{fig1}.

\section{Results}
\subsection{Characterizing actors spreading influence by TE networks using multiplex network measures}
\begin{figure*}[htbp]
    \centering
    \begin{subfigure}{0.27\linewidth}
        \centering
        \includegraphics[width=\linewidth]{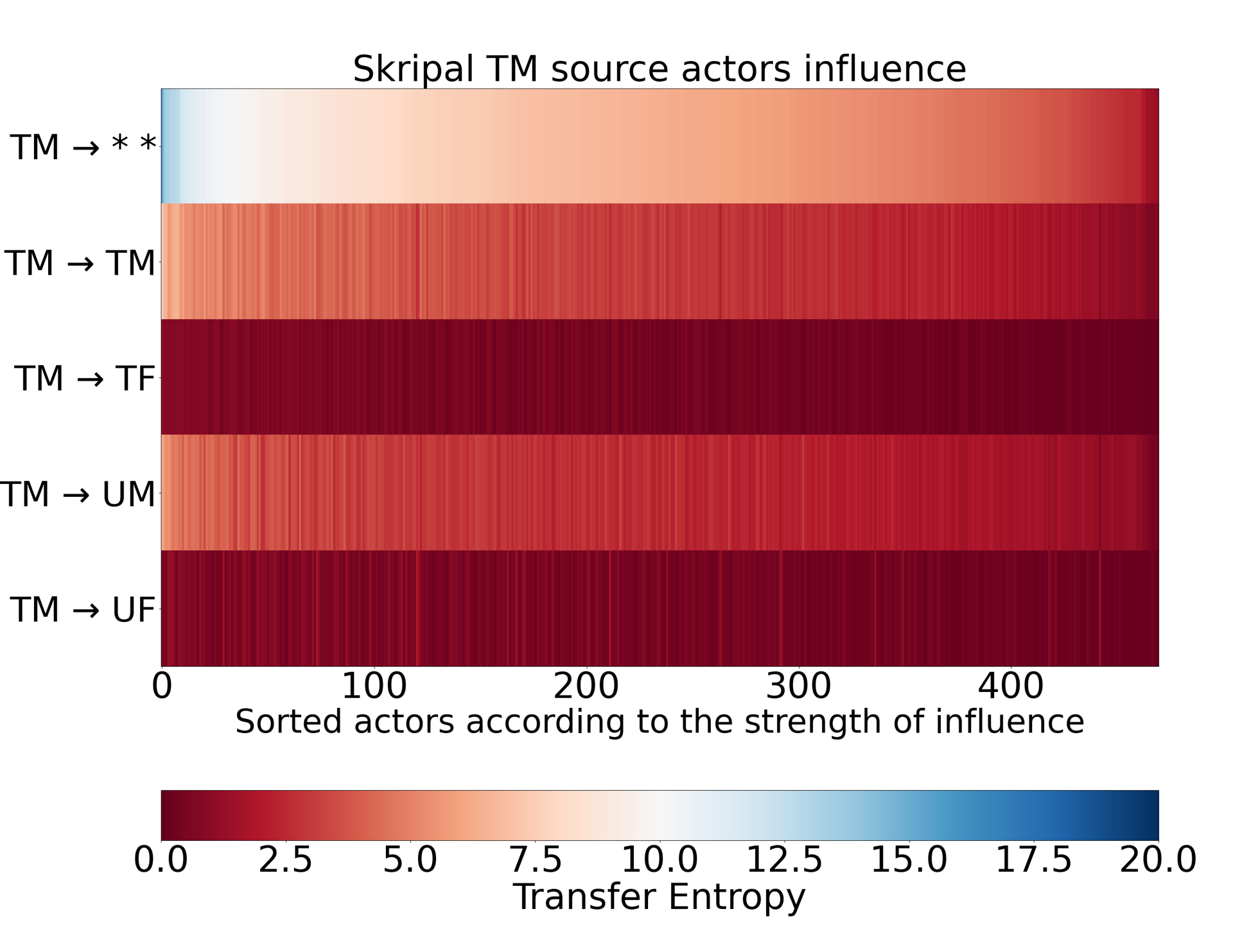}
    \end{subfigure}
    \hspace{5mm}
    \begin{subfigure}{0.27\linewidth}
        \centering
        \includegraphics[width=\linewidth]{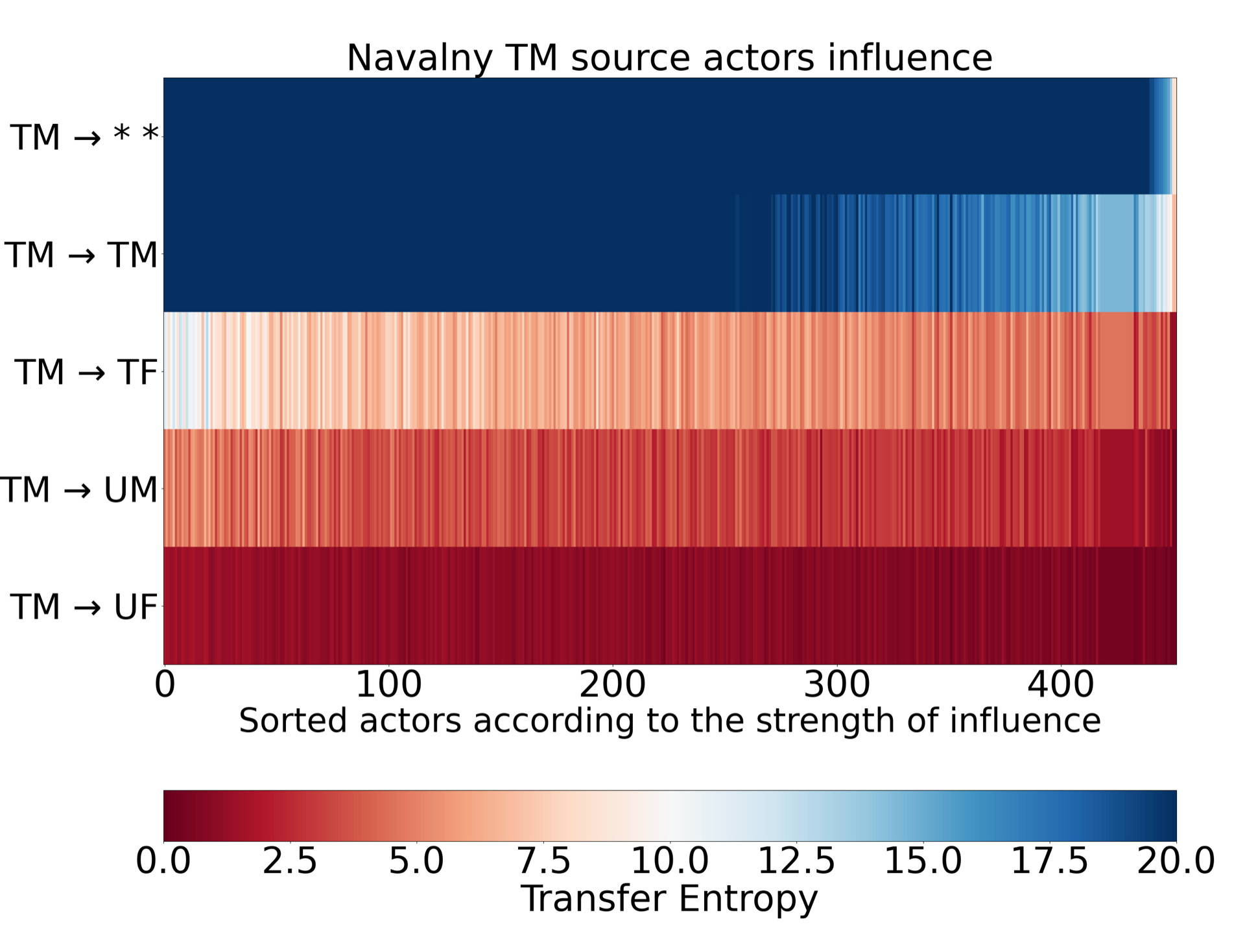}
    \end{subfigure}
    \hspace{5mm}
    \begin{subfigure}{0.27\linewidth}
        \centering
        \includegraphics[width=\linewidth]{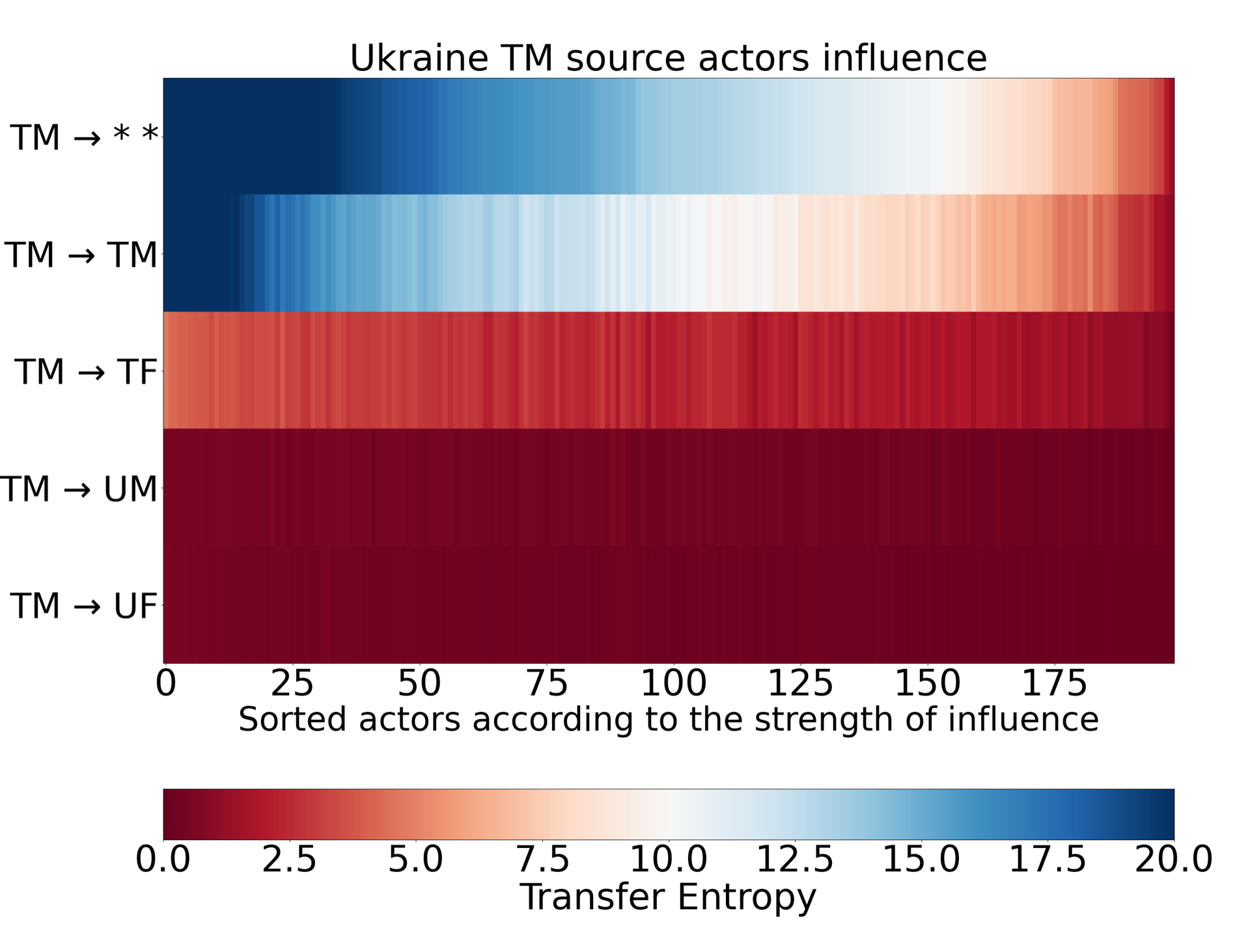}
    \end{subfigure}
    \begin{subfigure}{0.27\linewidth}
        \centering
        \includegraphics[width=\linewidth]{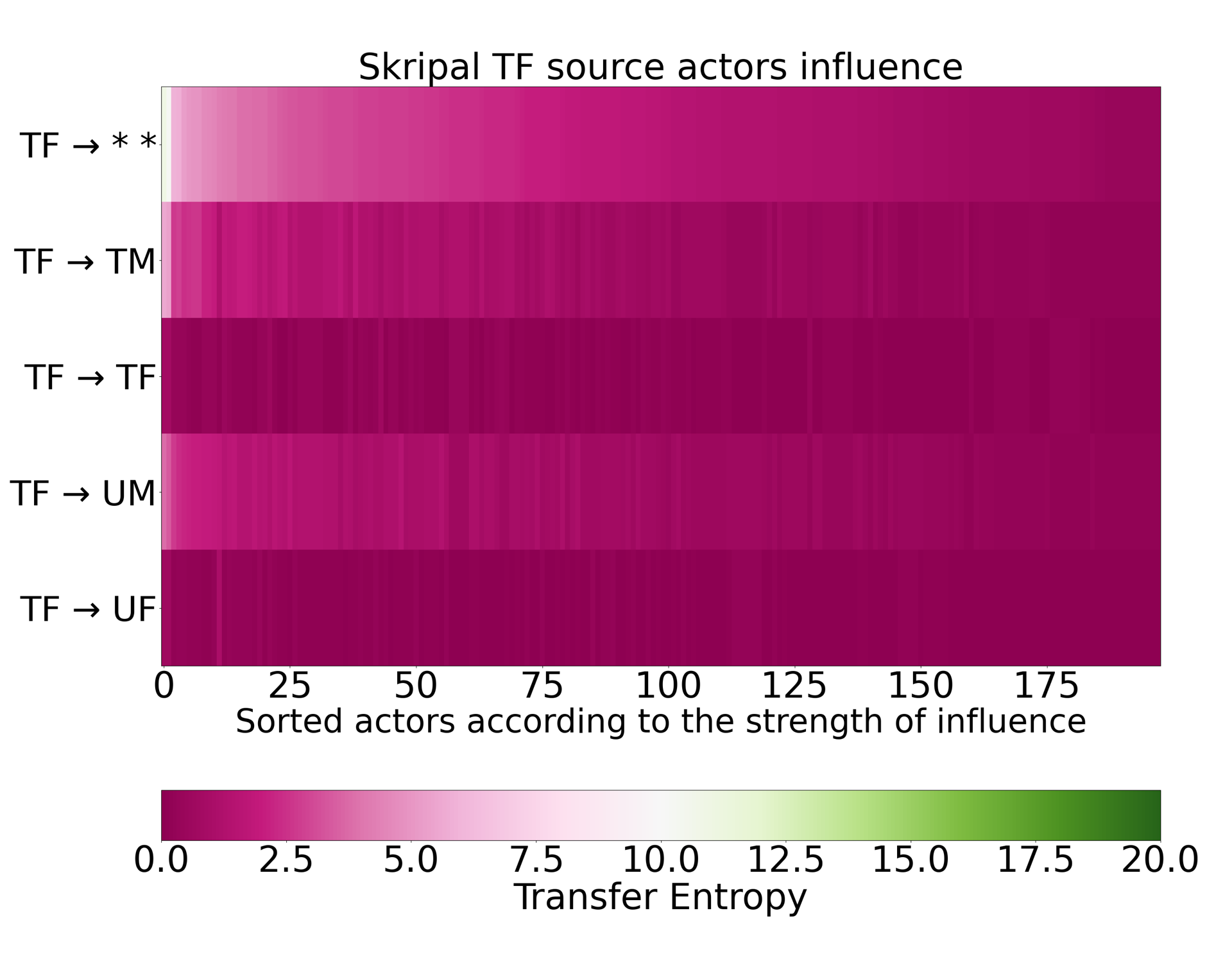}
    \end{subfigure}
    \hspace{5mm}
    \begin{subfigure}{0.27\linewidth}
        \centering
        \includegraphics[width=\linewidth]{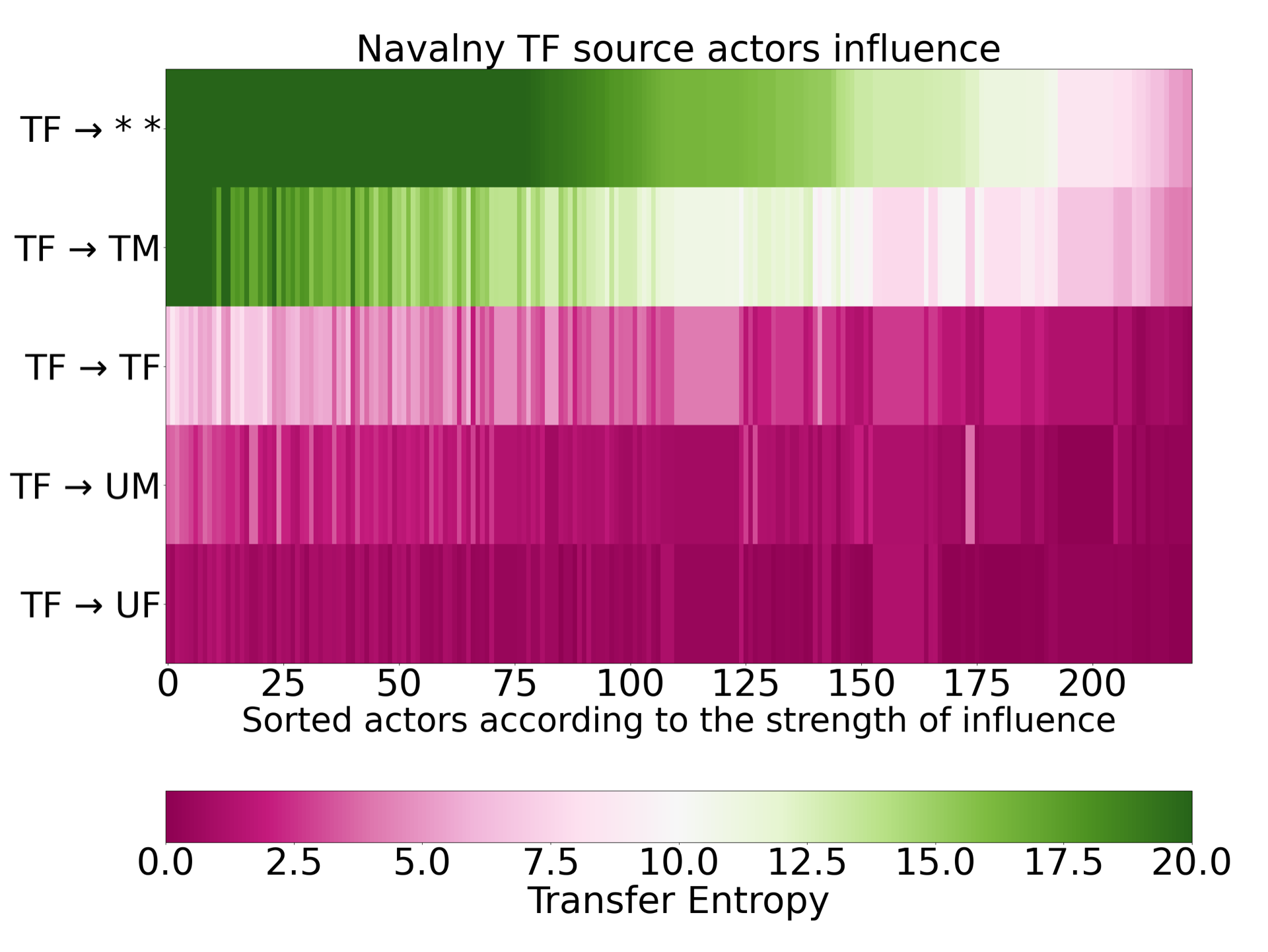}
    \end{subfigure}
    \hspace{5mm}
    \begin{subfigure}{0.27\linewidth}
        \centering
        \includegraphics[width=\linewidth]{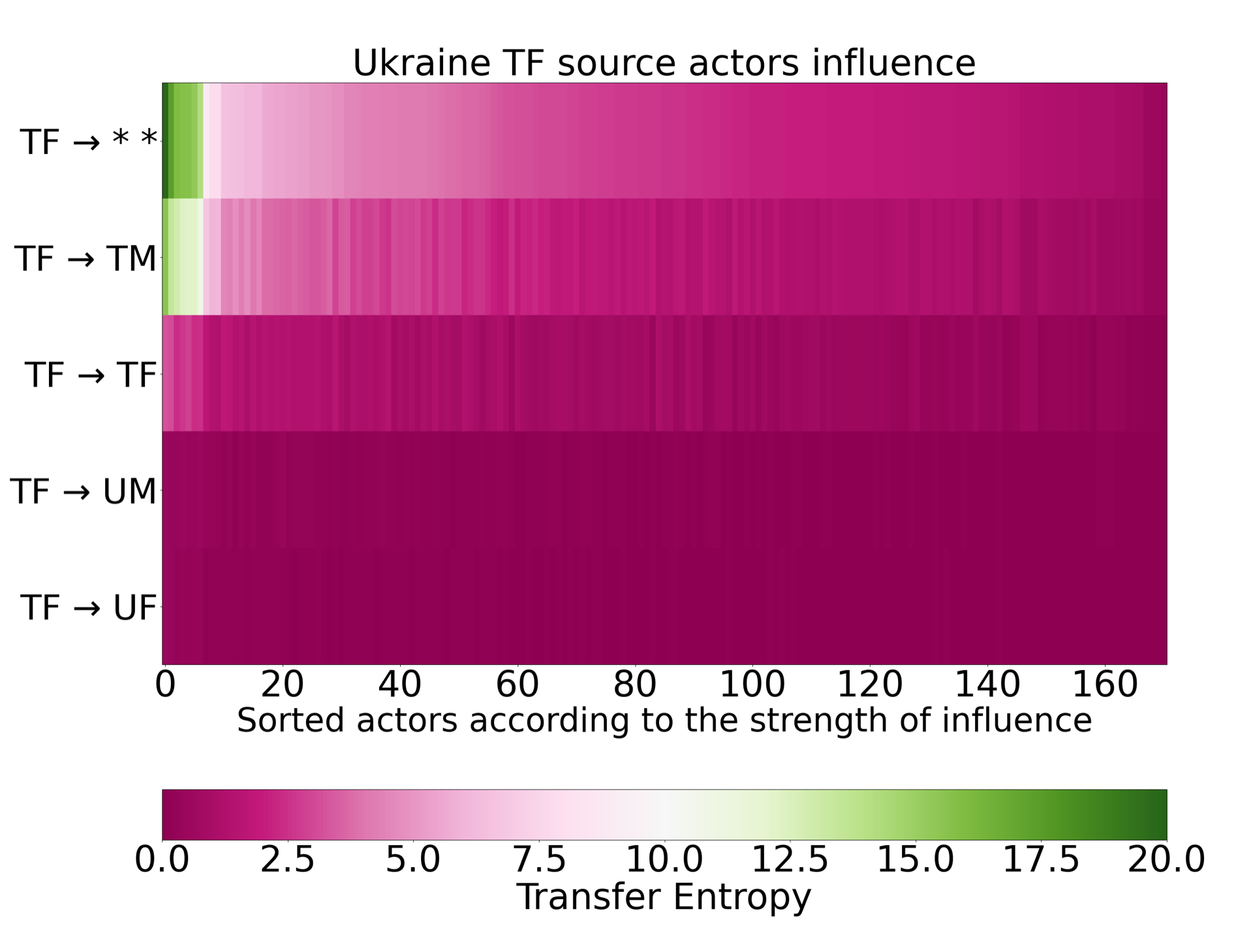}
    \end{subfigure}
    \begin{subfigure}{0.27\linewidth}
        \centering
        \includegraphics[width=\linewidth]{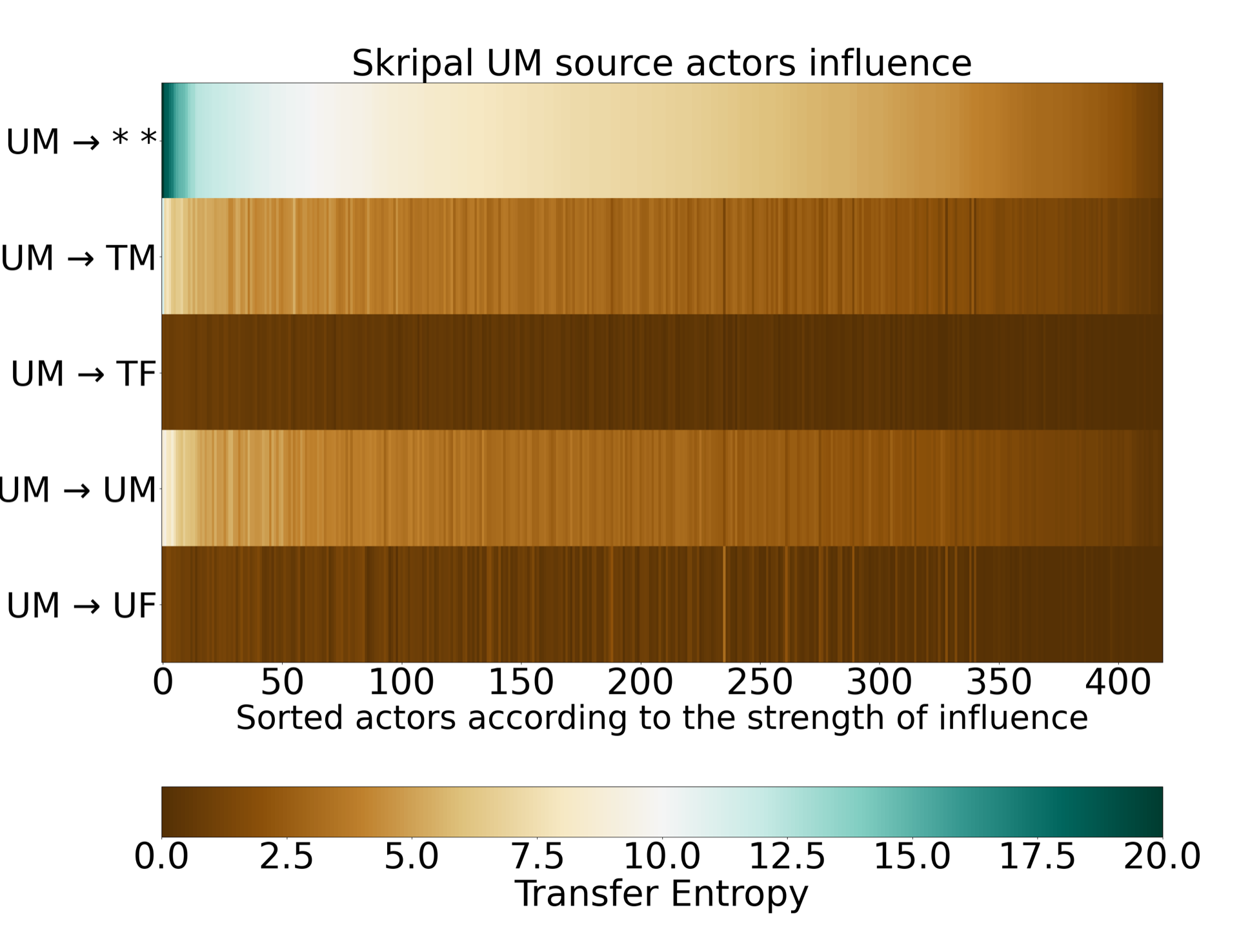}
    \end{subfigure}
    \hspace{5mm}
    \begin{subfigure}{0.27\linewidth}
        \centering
        \includegraphics[width=\linewidth]{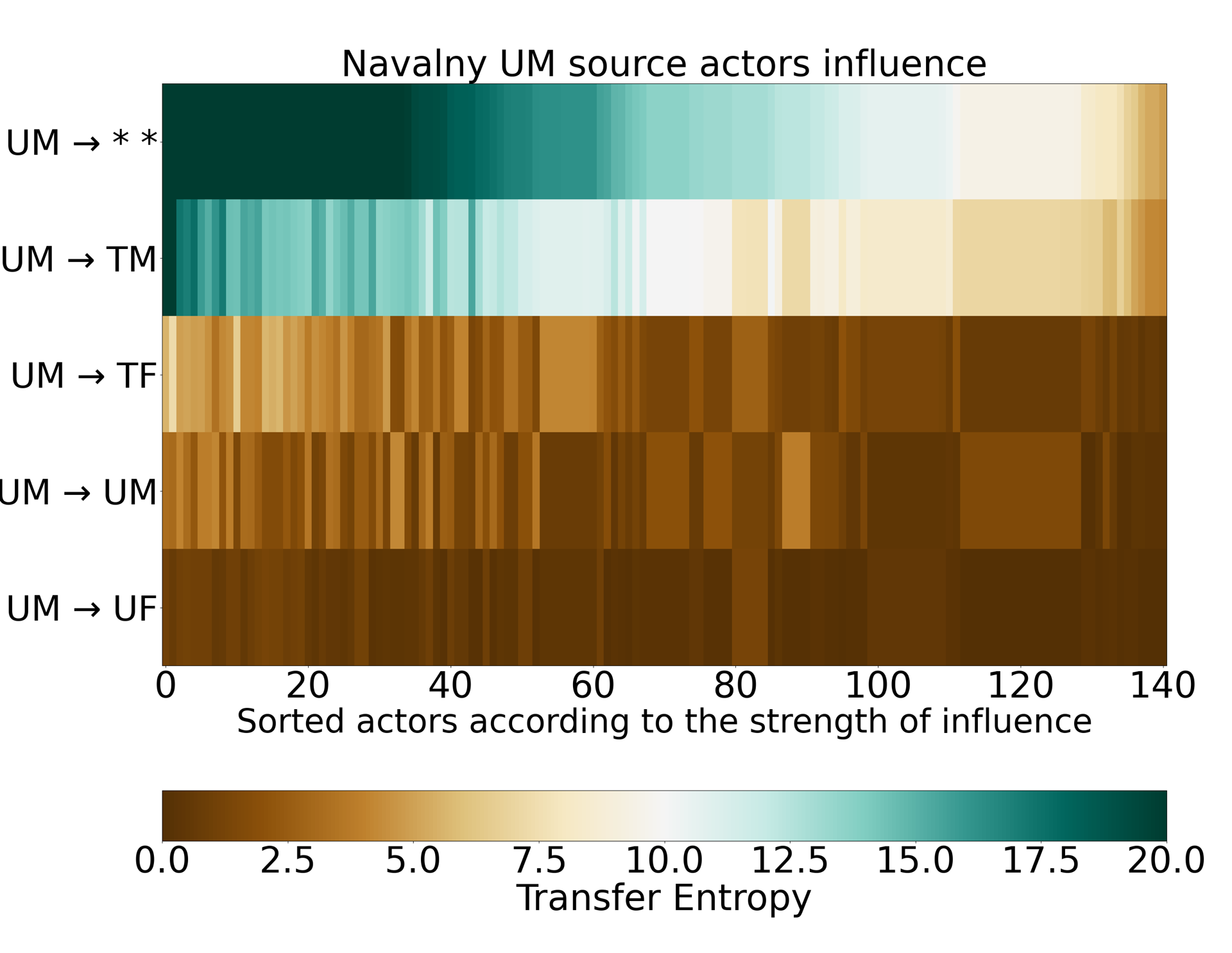}
    \end{subfigure}
    \hspace{5mm}
    \begin{subfigure}{0.27\linewidth}
        \centering
        \includegraphics[width=\linewidth]{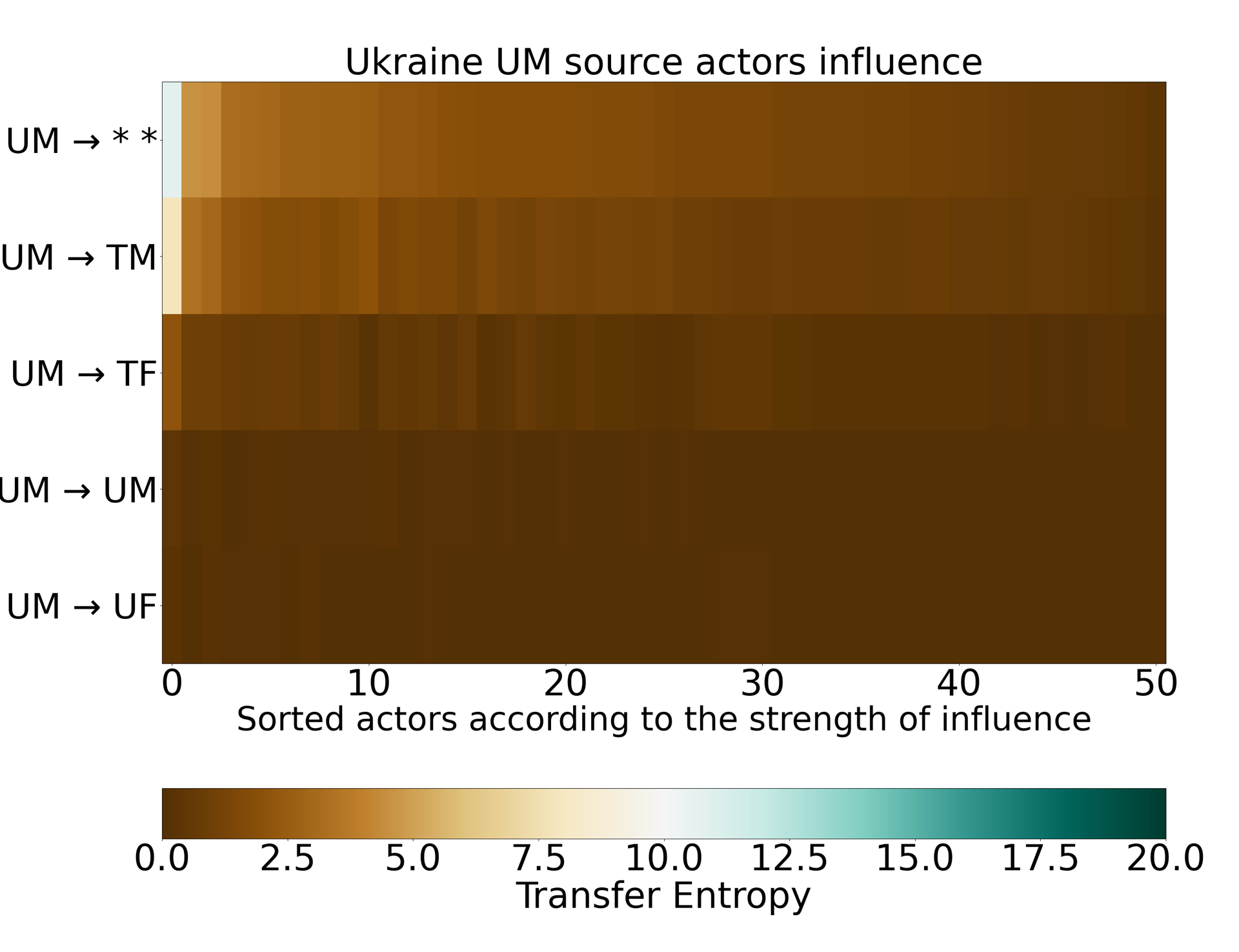}
    \end{subfigure}
    \begin{subfigure}{0.27\linewidth}
        \centering
        \includegraphics[width=\linewidth]{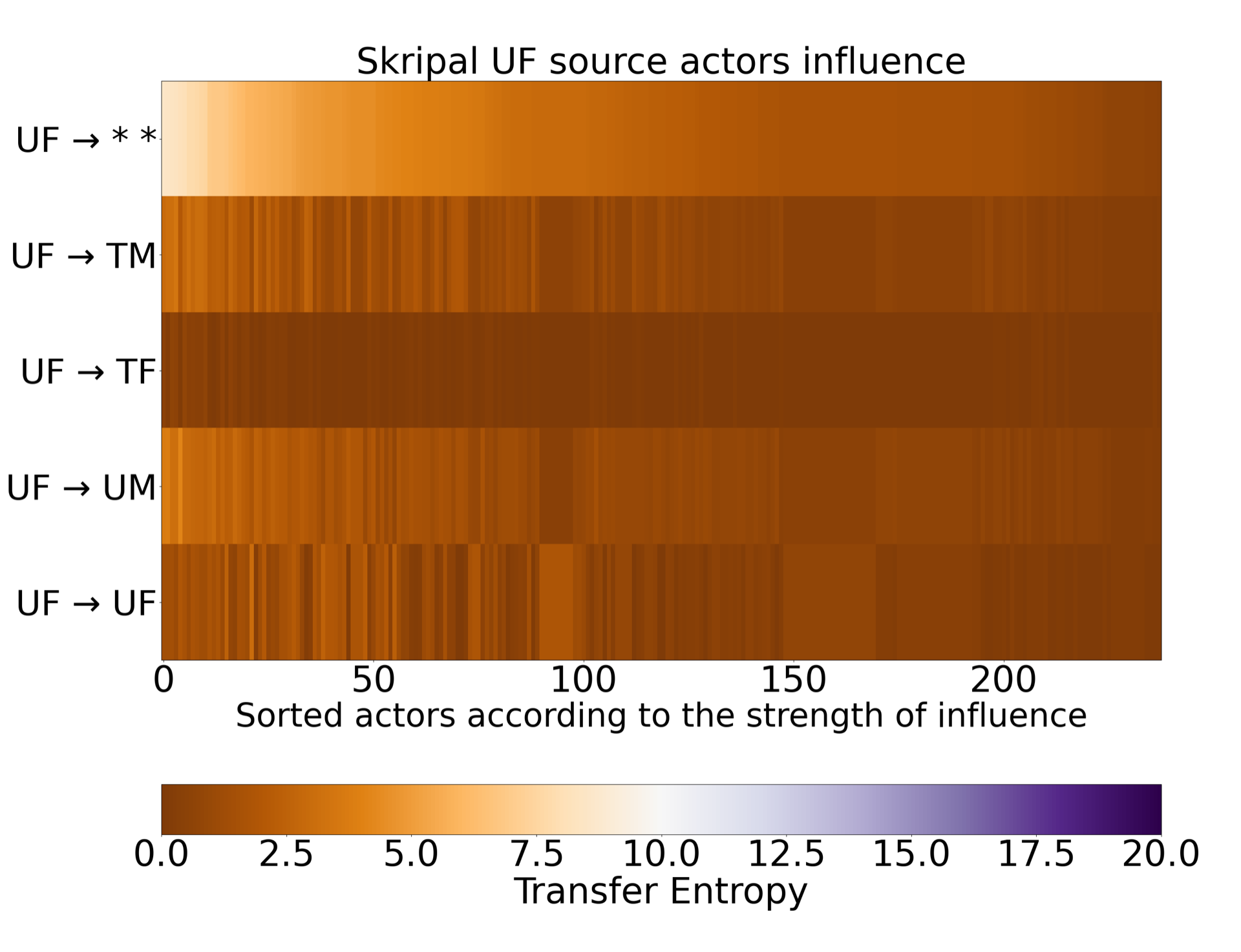}
        \caption{Skripal event}
    \end{subfigure}
    \hspace{5mm}
    \begin{subfigure}{0.27\linewidth}
        \centering
        \includegraphics[width=\linewidth]{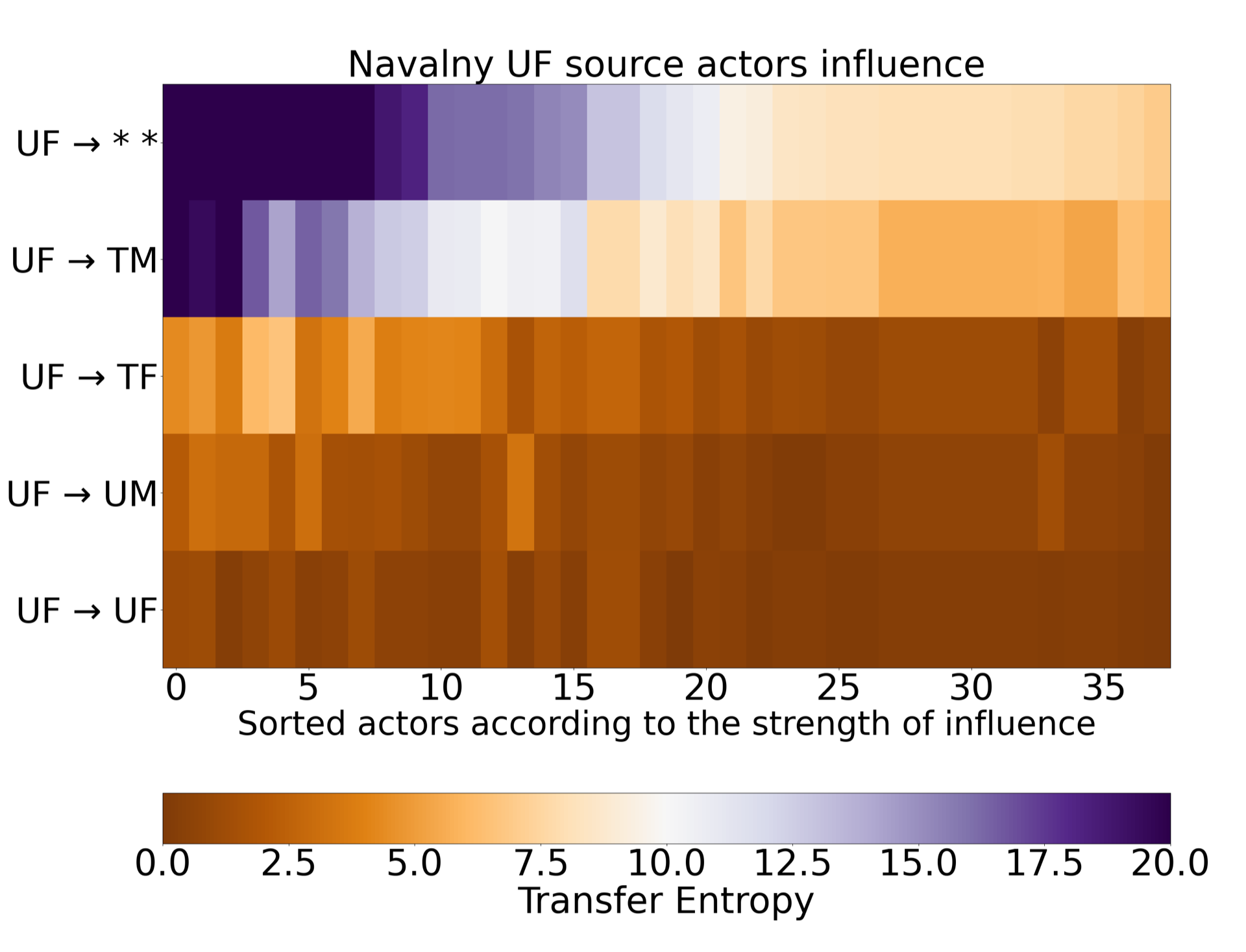}
        \caption{Navalny event}
    \end{subfigure}
    \hspace{5mm}
    \begin{subfigure}{0.27\linewidth}
        \centering
        \includegraphics[width=\linewidth]{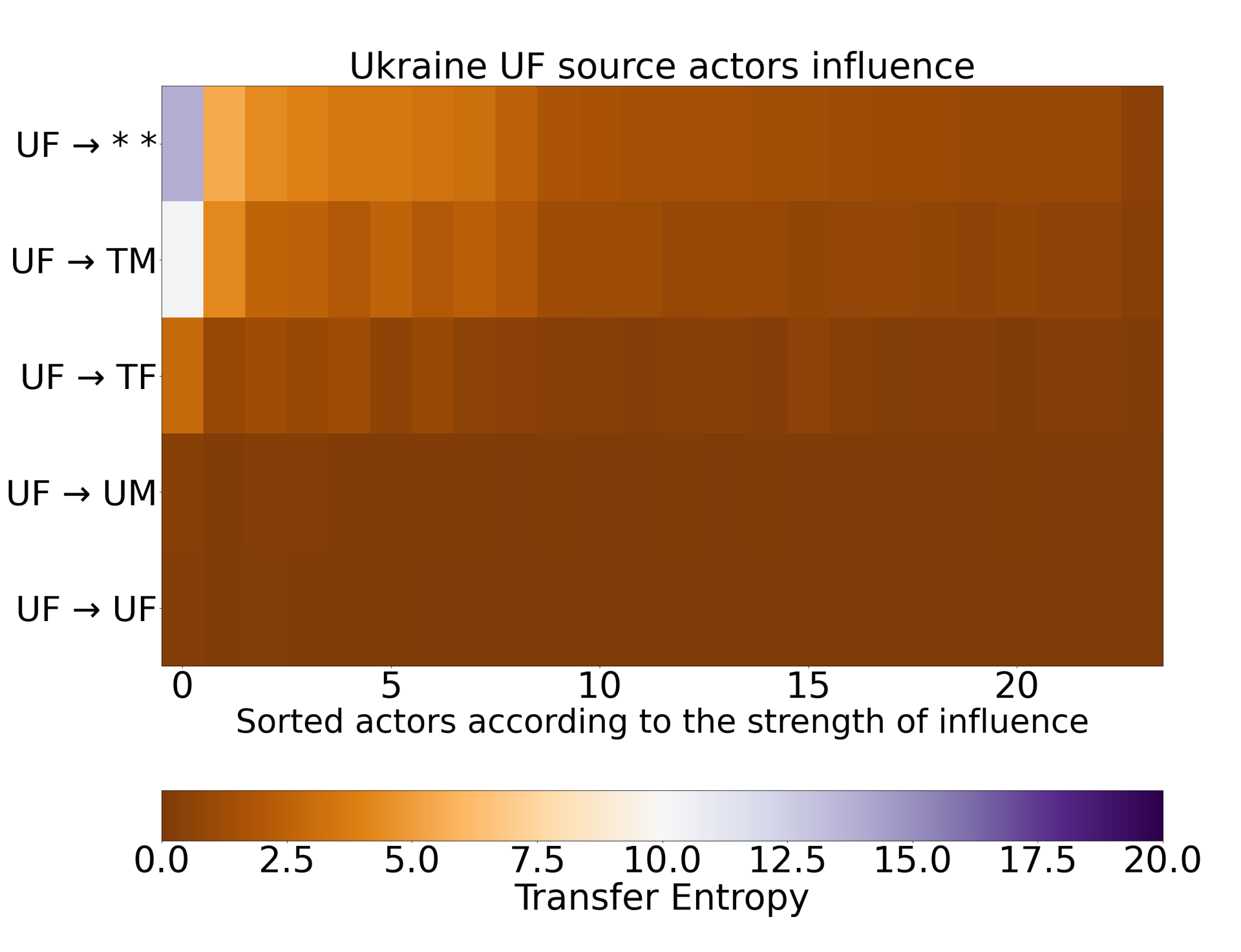}
        \caption{Ukraine event}
    \end{subfigure}
    \caption{Comparing the distribution of TM, TF, UM, and UF sources on four different type of targets in Skripal, Navalny, and Ukraine events}
    \label{fig3}
\end{figure*}

To employ multiplex network measures introduced in previous section and discern the influence distribution by actors, we generated four distinct multiplex networks for each influential actor type. Subsequently, actors were ranked based on the strength of influence exerted on their targets, measured by the total outdegree of a node in a multiplex network which is based on Transfer Entropy. Our objective is to investigate whether the strength of influence in a network has an effect on the distribution of influence to their targets. This methodology was applied to all Skripal, Navalny, and Ukraine events, and the results for each type of influential actor within their multiplex network are presented at \autoref{fig3}.

Comparing the two assassination events, the distribution of influence among Trustworthy Mainstream (TM) actors in the Skripal event was more uniform compared to the Navalny event. In the case of Navalny, the distribution of influence exerted by TM actors was not as uniform, although it did extend, to some degree, towards each type of target. Moreover, there was a significant activity in echo-chamber of TM actors influencing TM targets in the Navalny event. However, during the Ukraine event, TM actors predominantly influenced targets who were tweeting from trustworthy news sources. This distinction is apparent in their participant coefficient charts. In the Skripal event, a majority of TM actors exhibit a high participant coefficient, indicating a more diversified distribution of influence among different targets with varying influence types. Conversely, in the Ukraine event, a low participant coefficient is observed when the amount of TE was high, suggesting that TM actors with higher levels of influence engaged in more focused operations compared to those with lower influence. In the Navalny event, given the less uniform distribution of influence by TM actors, it is observable that most were aligned with the green line, indicating a predominant influence over two layers. his pattern was almost identical among other types of influential actors. A noteworthy observation in the Ukraine incident was the absence of influence directed from actors towards UM and UF targets. This pattern of behavior was distinct from what was observed in both the Skripal and Navalny incidents which had similar nature, namely, the death of an individual.

\begin{figure}[htbp]
    \centering
    \begin{subfigure}{0.32\linewidth}
        \centering
        \includegraphics[width=\linewidth]{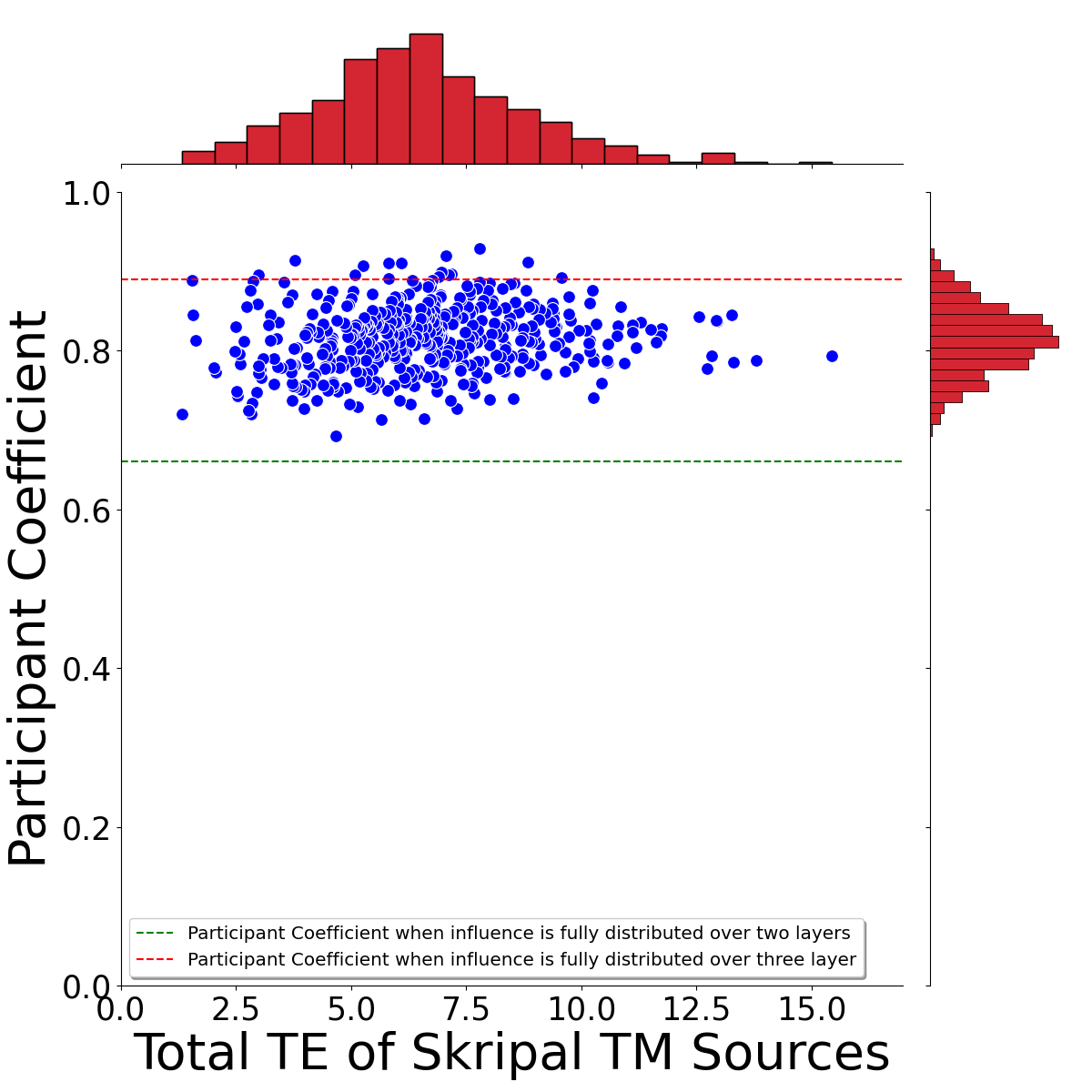}
    \end{subfigure}
    \hfill
    \begin{subfigure}{0.32\linewidth}
        \centering
        \includegraphics[width=\linewidth]{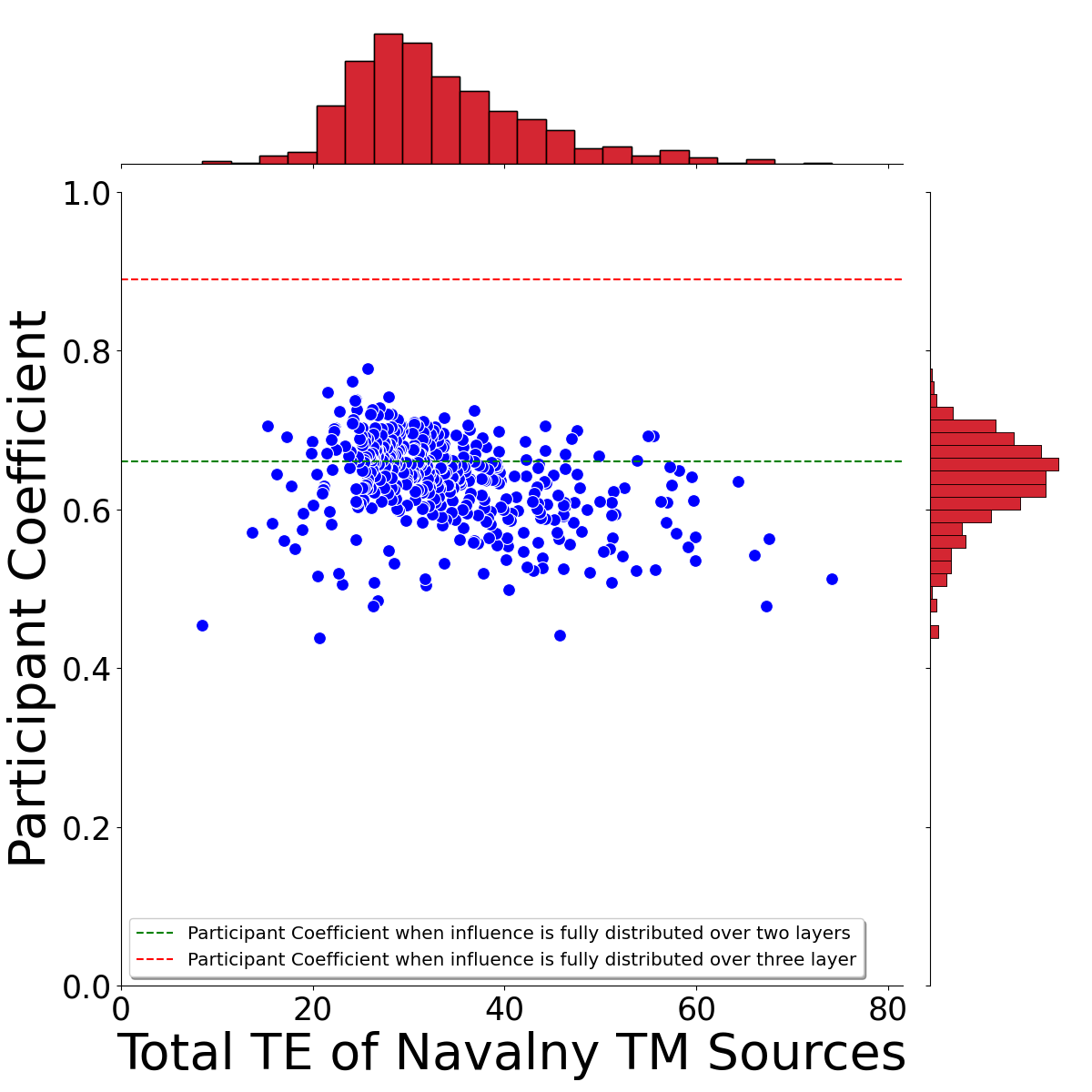}
    \end{subfigure}
    \hfill
    \begin{subfigure}{0.32\linewidth}
        \centering
        \includegraphics[width=\linewidth]{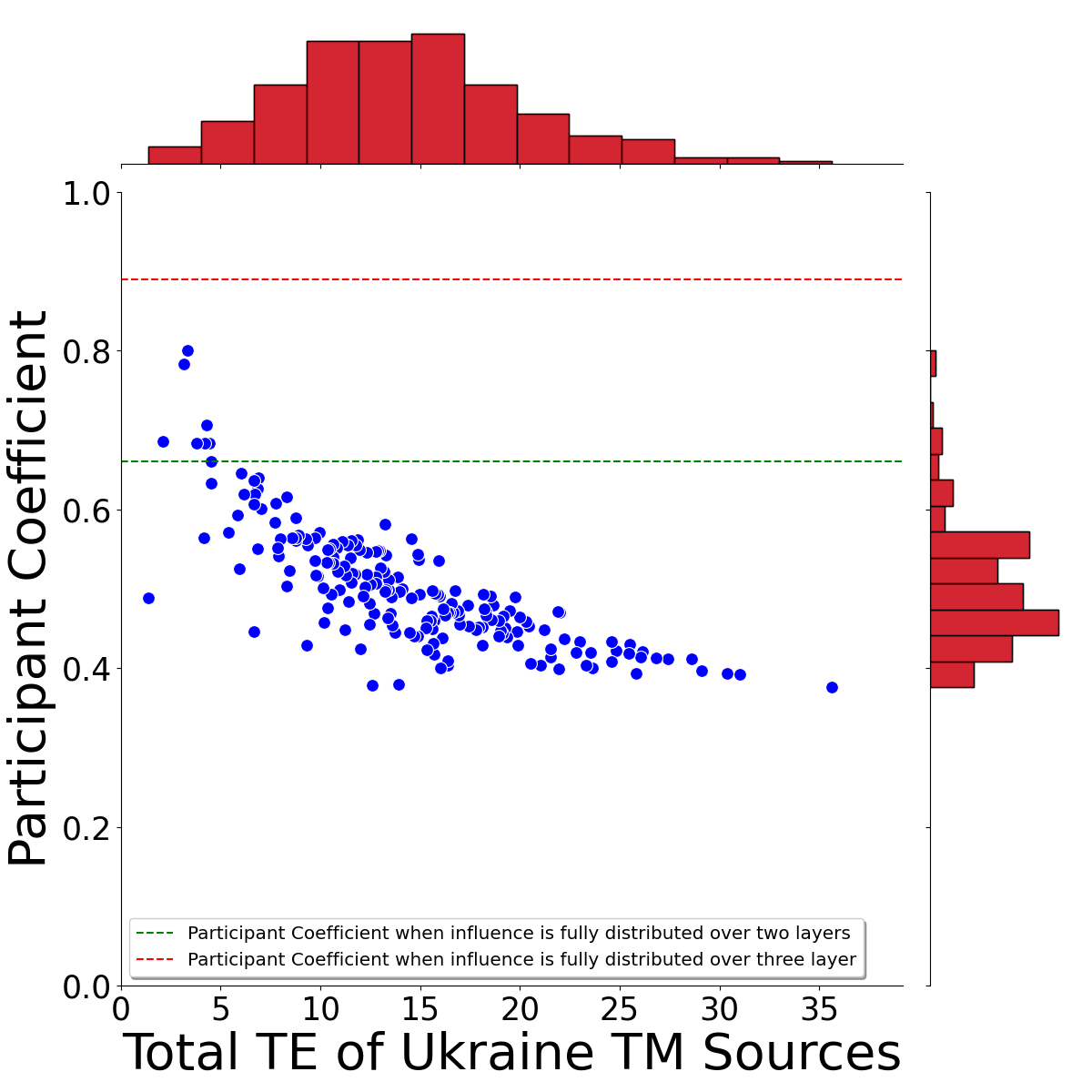}
    \end{subfigure}
    \begin{subfigure}{0.32\linewidth}
        \centering
        \includegraphics[width=\linewidth]{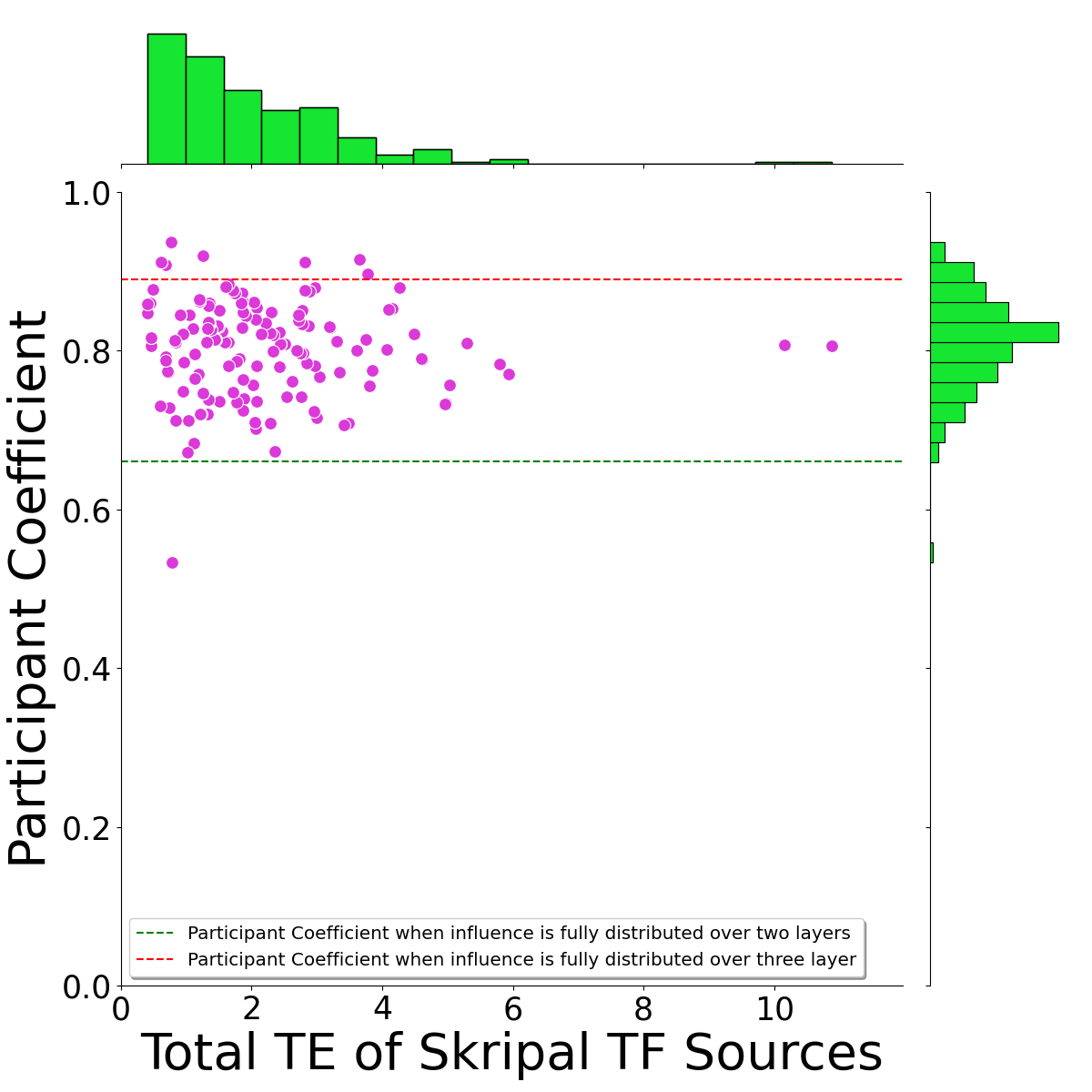}
    \end{subfigure}
    \hfill
    \begin{subfigure}{0.32\linewidth}
        \centering
        \includegraphics[width=\linewidth]{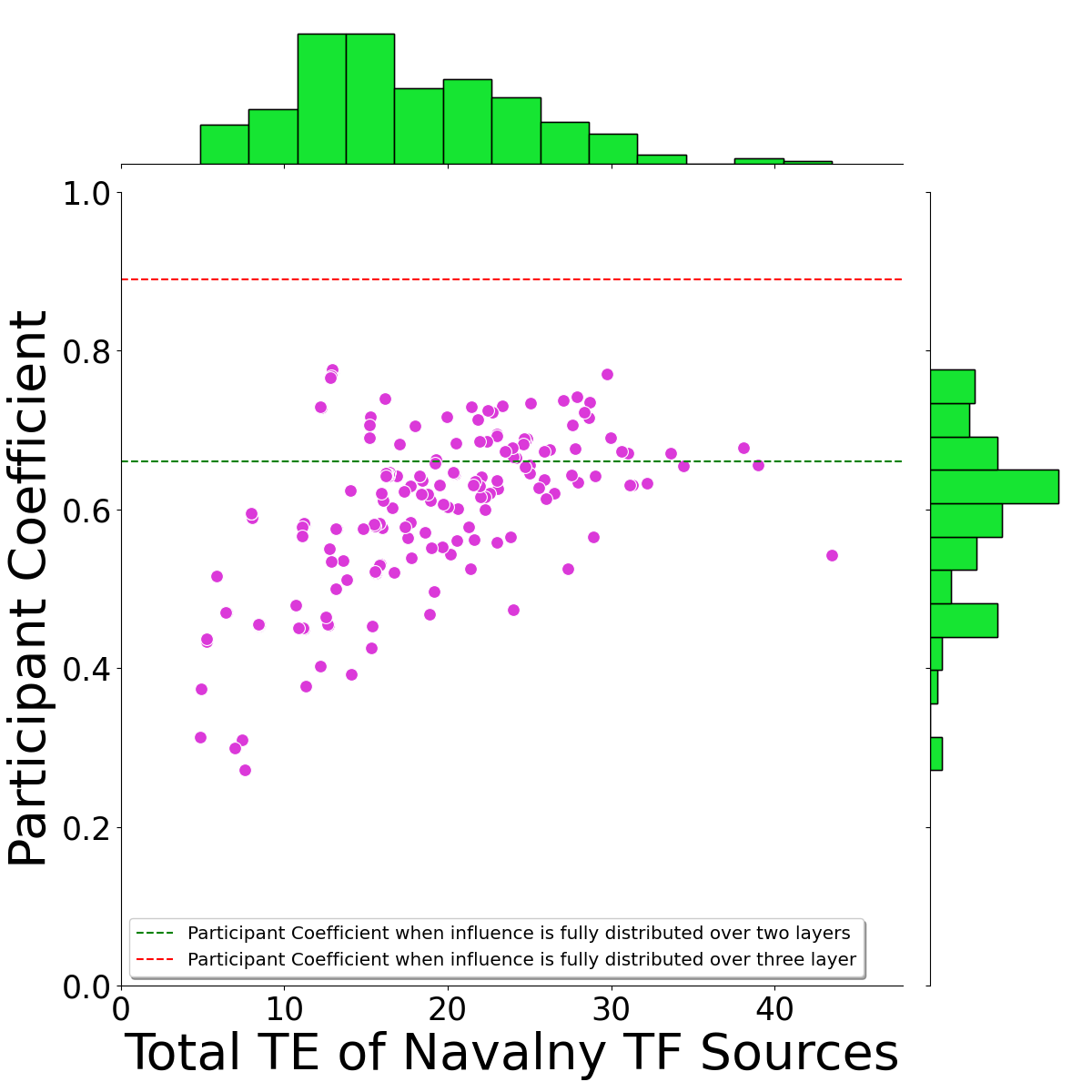}
    \end{subfigure}
    \hfill
    \begin{subfigure}{0.32\linewidth}
        \centering
        \includegraphics[width=\linewidth]{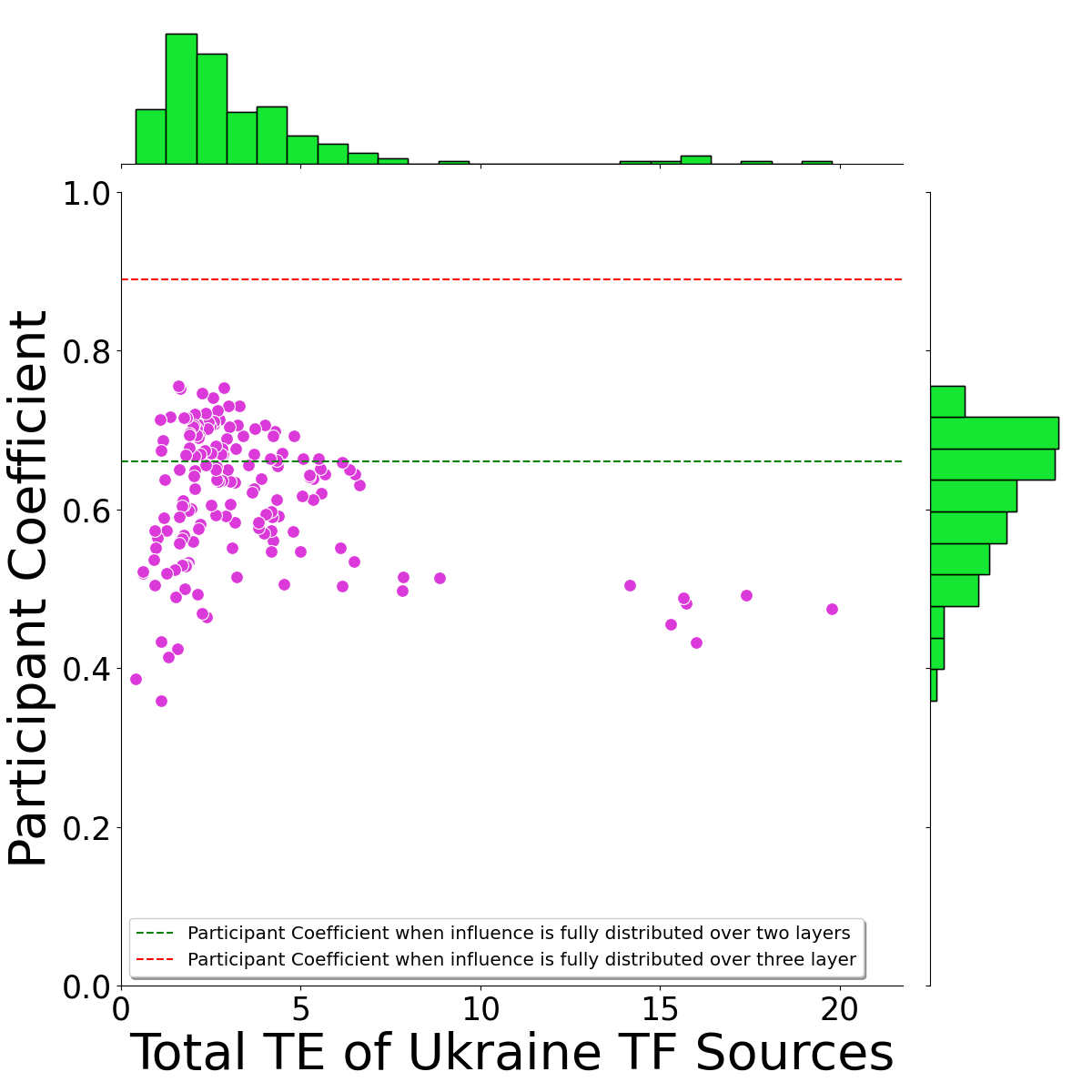}
    \end{subfigure}
    \begin{subfigure}{0.32\linewidth}
        \centering
        \includegraphics[width=\linewidth]{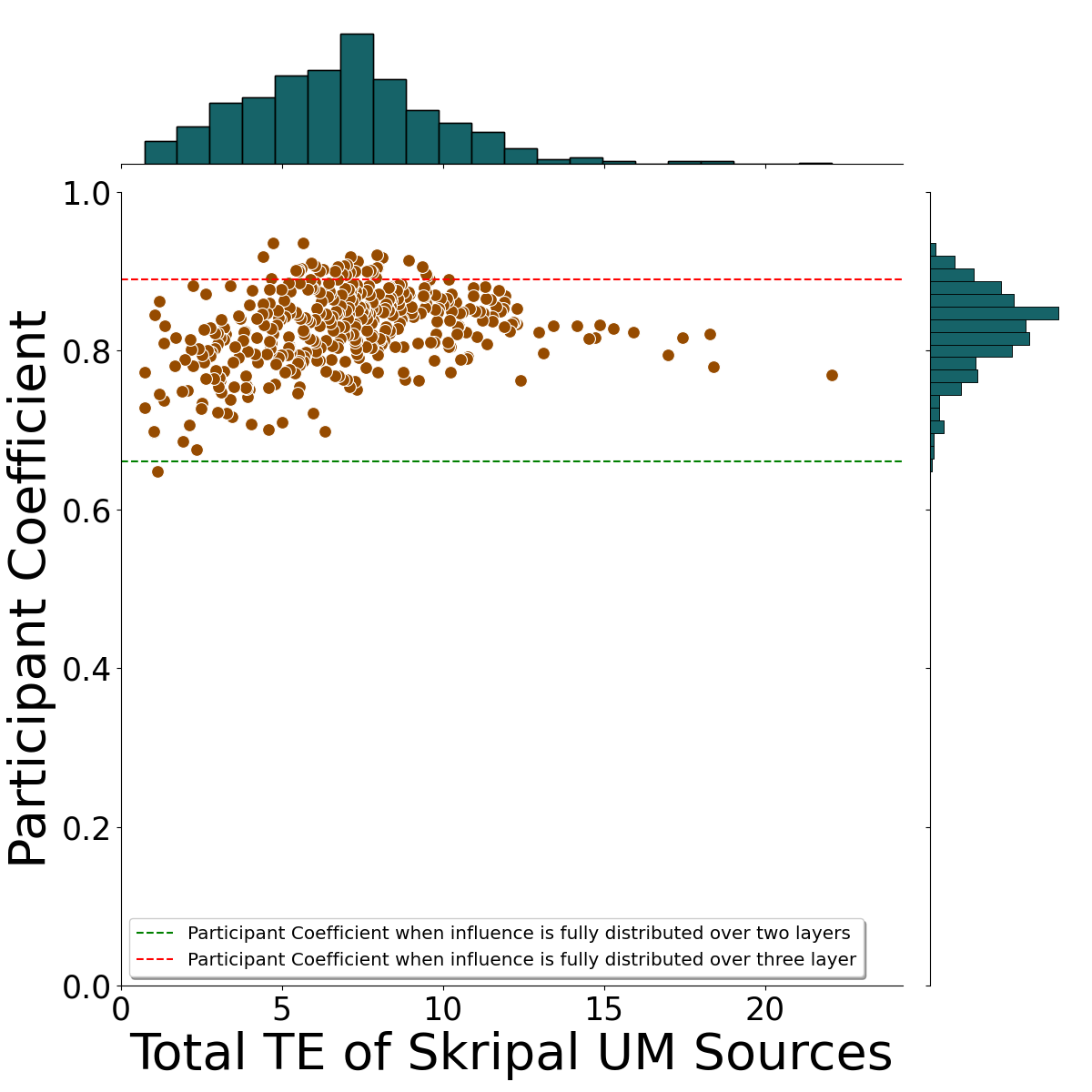}
    \end{subfigure}
    \hfill
    \begin{subfigure}{0.32\linewidth}
        \centering
        \includegraphics[width=\linewidth]{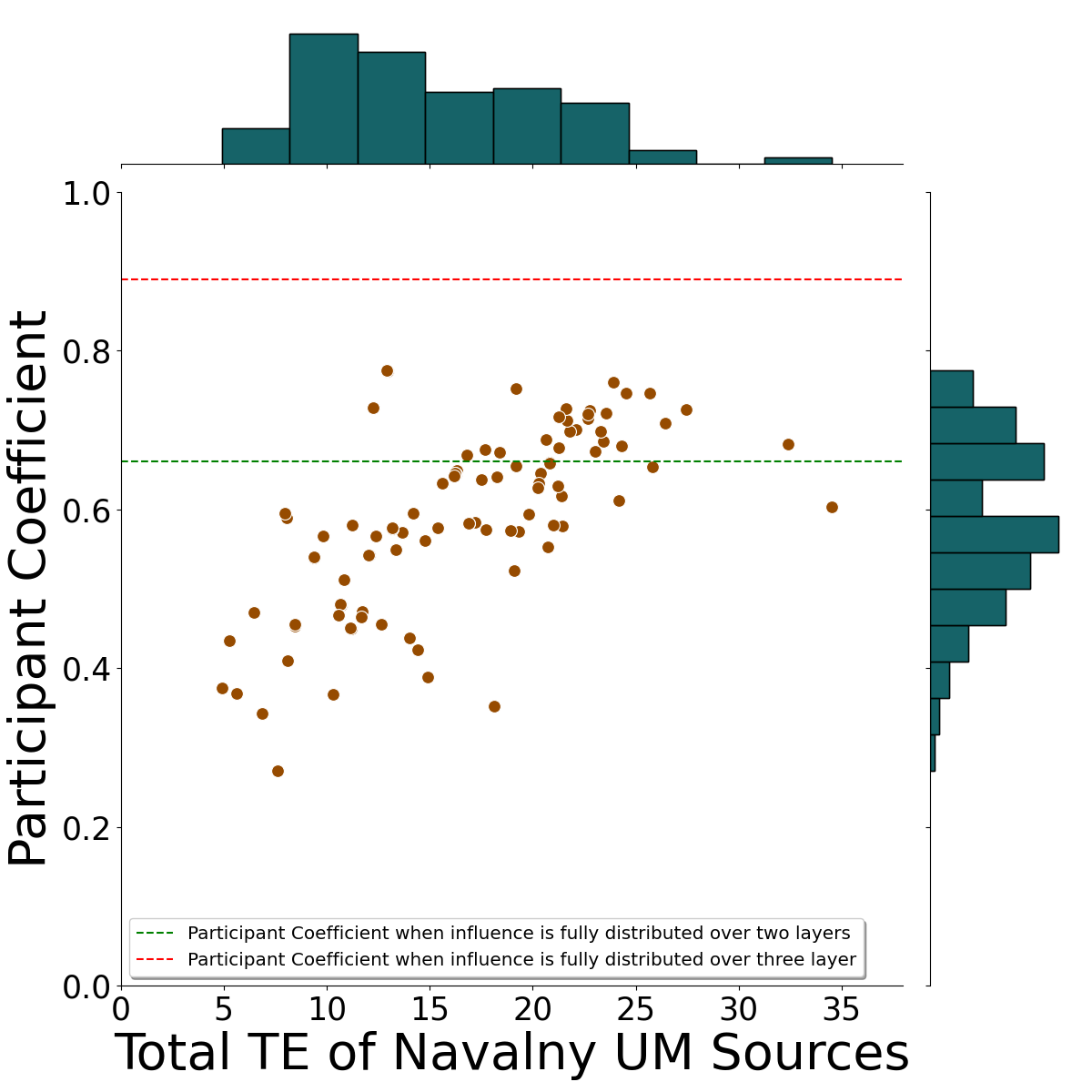}
    \end{subfigure}
    \hfill
    \begin{subfigure}{0.32\linewidth}
        \centering
        \includegraphics[width=\linewidth]{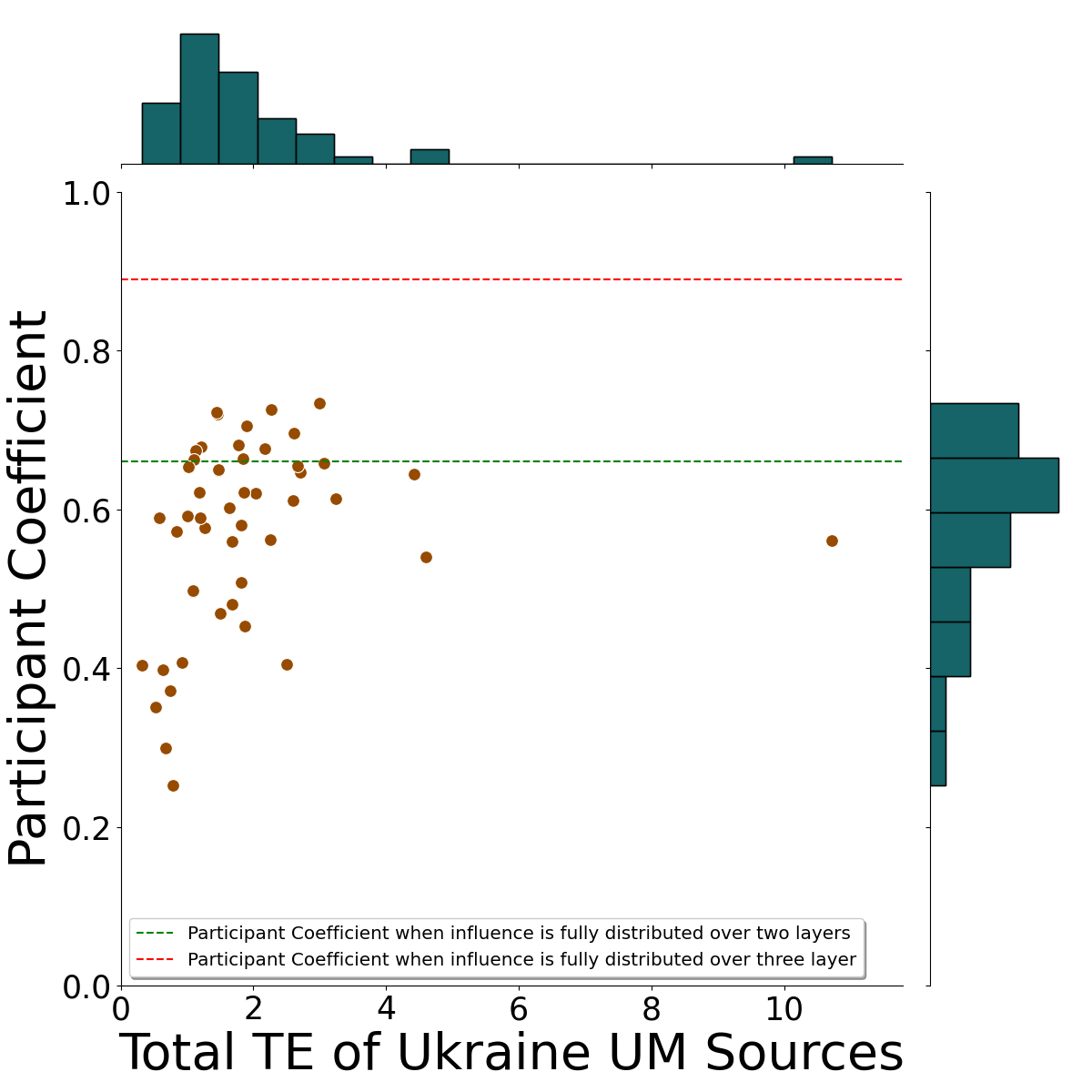}
    \end{subfigure}
    \begin{subfigure}{0.32\linewidth}
        \centering
        \includegraphics[width=\linewidth]{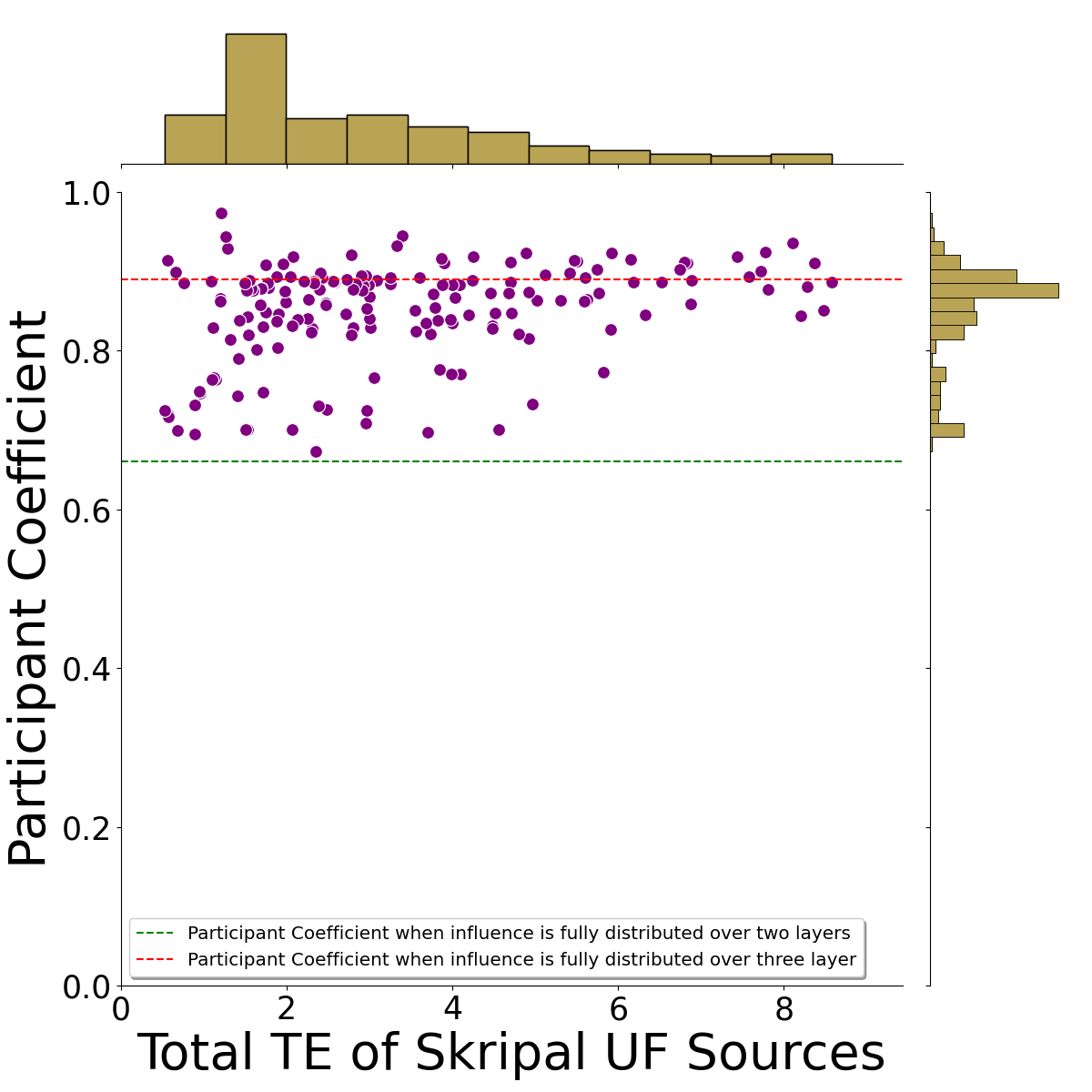}
        \caption{Skripal event}
    \end{subfigure}
    \hfill
    \begin{subfigure}{0.32\linewidth}
        \centering
        \includegraphics[width=\linewidth]{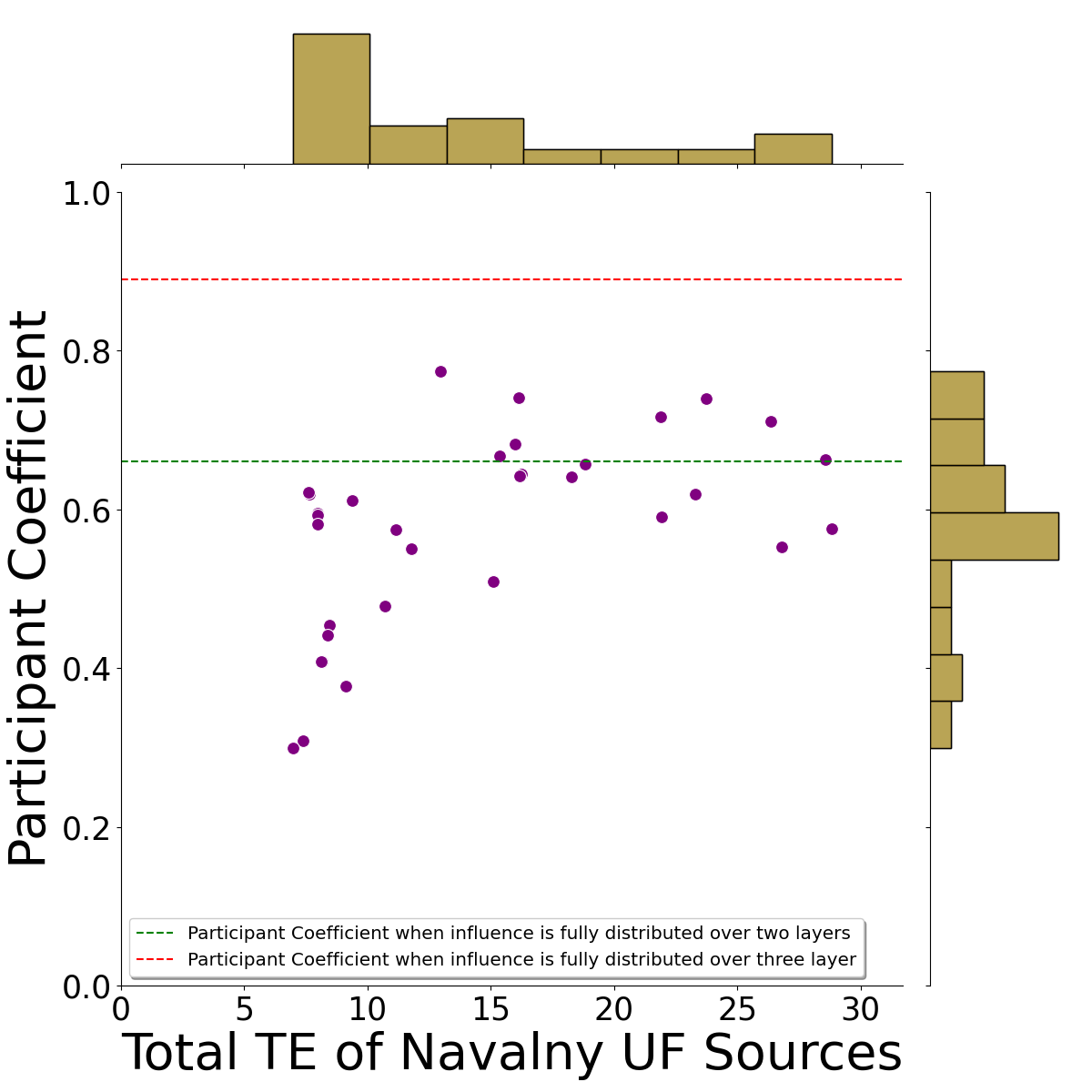}
        \caption{Navalny event}
    \end{subfigure}
    \hfill
    \begin{subfigure}{0.32\linewidth}
        \centering
        \includegraphics[width=\linewidth]{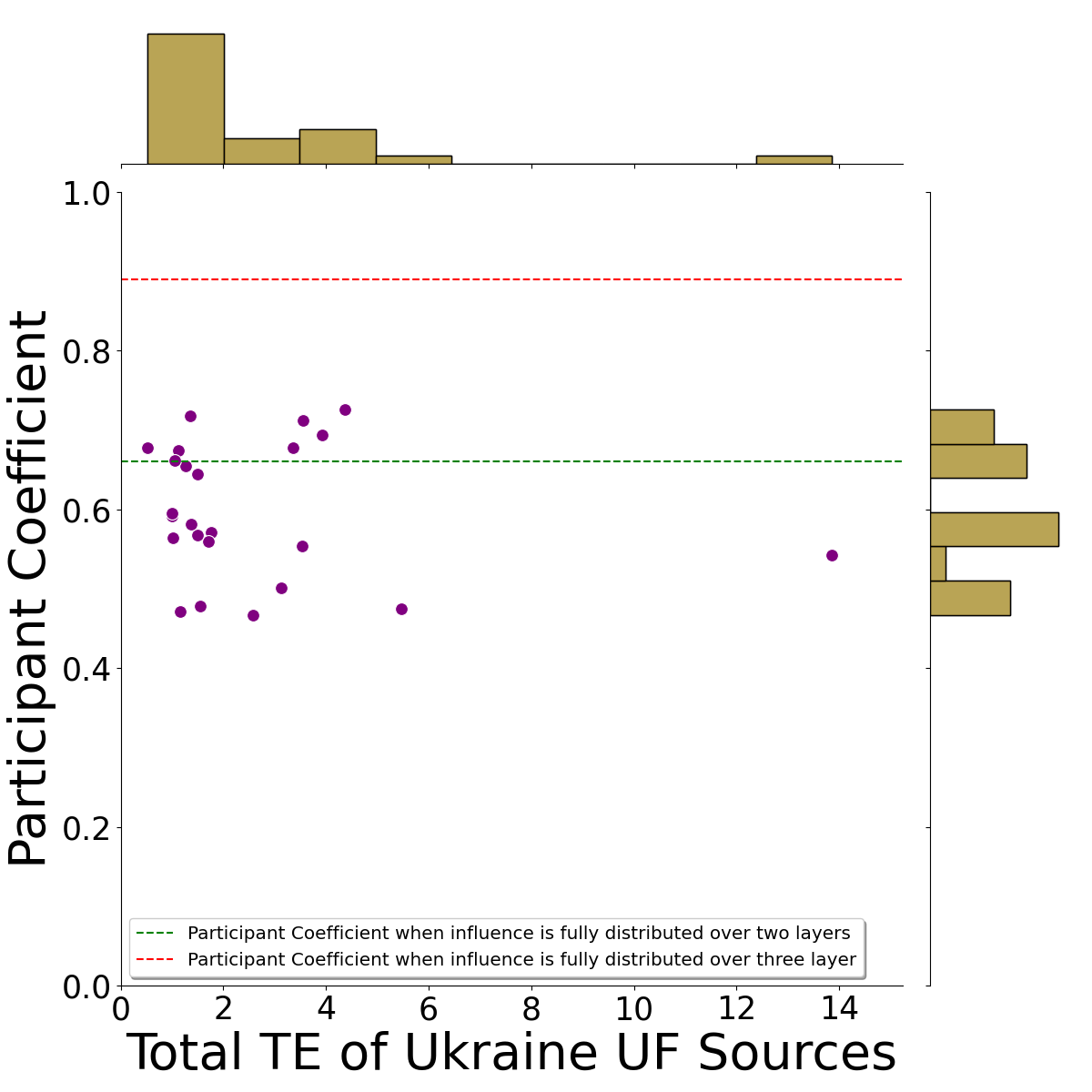}
        \caption{Ukraine event}
    \end{subfigure}
    \caption{The participant coefficient of TM, TF, UM, and UF actors in their corresponding multiplex networks in Skripal, Navalny, and Ukraine events \\
    Note: The scales on the X-axis are adjusted according to the highest influence in the multiplex network. A single scale was not used in order to highlight the diversity of influence and to distinguish the distributions.}
    \label{fig4}
\end{figure}

Examining the participant coefficient plots in \autoref{fig4} reveals that during the Skripal incident, the distribution of influence for almost all actors is predominantly centered around three layers of influence, as indicated by the red dotted lines. Conversely, in the Navalny incident, with the exception of TM actors, who are primarily concentrated around two layers of influence denoted by green dotted line, the behavior of actors in other categories exhibits notable variance. In the TF, UM, and UF multiplex networks, actors possessing lower levels of influence demonstrate a more focused approach towards their targets compared to those wielding a higher degree of influence. This trend is starkly reversed in the context of the Ukraine incident, where actors with significant influence exhibit a more concentrated focus towards their targets, in contrast to less influential actors, whose influence is dispersed across multiple layers.    

\subsection{Pairwise co-occurrence comparison}
\begin{figure*}[htbp]
    \centering
    \begin{subfigure}{0.32\linewidth}
        \centering
        \includegraphics[width=\linewidth]{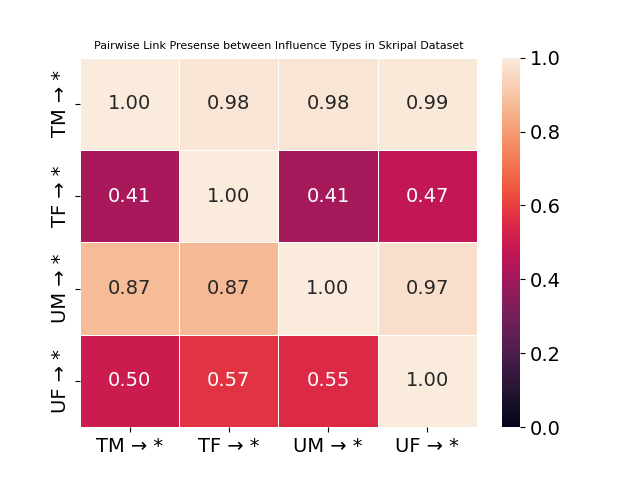}
        \caption{Skripal event}
    \end{subfigure}
    \hfill
    \begin{subfigure}{0.32\linewidth}
        \centering
        \includegraphics[width=\linewidth]{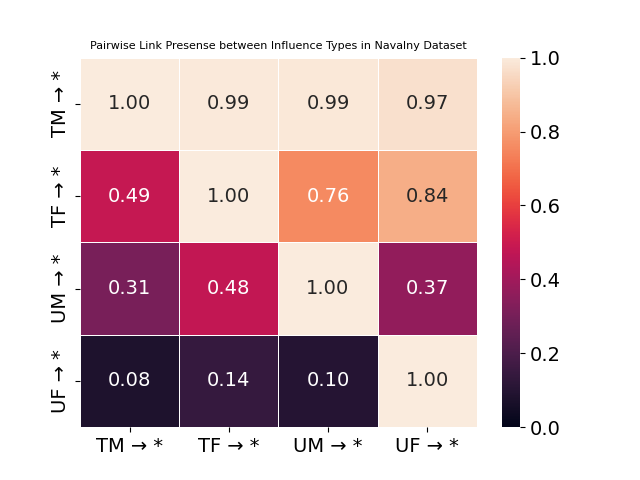}
        \caption{Navalny event}
    \end{subfigure}
    \hfill
    \begin{subfigure}{0.32\linewidth}
        \centering
        \includegraphics[width=\linewidth]{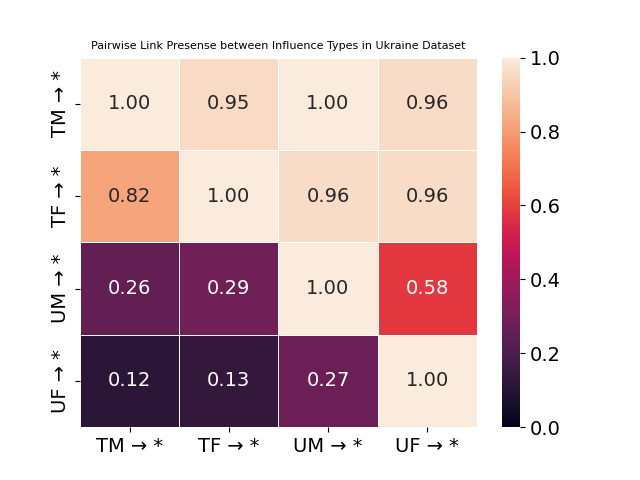}
        \caption{Ukraine event}
    \end{subfigure}
    \caption{The pairwise co-occurrence of influential actors within a multiplex network being active in other types of influential multiplex networks \\
    The asterisk (*) in the heatmaps signifies that influence extended to all types of targets (TM, TF, UM, UF)}
    \label{fig5}
\end{figure*}

In the next step, we conducted pairwise co-occurrence analyses for different actors. Our objective was to explore the likelihood that an actor, for instance, initially tweeting from trustworthy mainstream news sources and influencing others, would subsequently engage with other news types and exert influence on their respective targets. The results represented in heatmaps in \autoref{fig5}.

Upon examining the heatmaps, it can be inferred that all three events exhibit a similar pattern to a significant extent. Notably, the percentages above the diagonal are high, whereas those below the diagonal are low. Specifically, the transition from trustworthy mainstream (TM) and trustworthy fringe (TF) actors to untrustworthy mainstream (UM) and untrustworthy fringe (UF) actors suggests that individuals engaging with trustworthy news sources are more likely to diversify their news consumption, potentially including untrustworthy sources. For the Skripal, Navalny, and Ukraine cases, 98\%, 99\%, and 100\% of actors utilizing TM news sources also engaged with UM sources, respectively. Conversely, only 87\%, 31\%, and 26\% of actors associated with the Skripal, Navalny, and Ukraine events, respectively, who initially used UM news sources, were observed to utilize TM sources. This indicates a diminished propensity for actors primarily influencing with untrustworthy news sources to transition to utilizing trustworthy sources, underscoring a more targeted and focused influence strategy among those relying on untrustworthy news. In contrast, actors who disseminate trustworthy news demonstrate a willingness to explore a broader range of media and platforms, highlighting their diverse information consumption patterns.

\section{Conclusion}
In this paper, we explored the propagation of influence within the social network platform X by examining the behavior of actors who utilize four distinct classes of news articles. Diverging from traditional methodologies that rely on traditional methods such as nodes' degrees and centrality measures to assess influence and identify influential actors, we employed a non-parametric approach named Transfer Entropy. This method has the capability to uncover hidden influencers not detectable by parametric and conventional techniques. To analyze the behavior of these actors, we adopted a novel metrics based on multiplex networks. Through the construction of these networks and the examination of influence distribution, we observed that in assassination-related scenarios, especially in the Navalny case, trustworthy actors were targeted by untrustworthy actors. Conversely, in the context of the Ukraine war, the predominant influence was exerted by actors utilizing trustworthy news sources and the absence of untrustworthy sources was obvious. Additionally, looking at the participant coefficient, we discovered that the behavior of actors utilizing trustworthy sources in assassination scenarios were similar to one another across different strengths of influence. However, in the Ukraine war scenario, particularly within the Trustworthy Mainstream (TM) multiplex network, the influence exerted by highly influential actors was more concentrated through a single network, rather than being dispersed across multiple layers of networks. A significant finding is the tendency of actors using trustworthy news sources to also employ untrustworthy news articles, indicating a propensity for these actors to utilize a diverse range of news articles in terms of credibility. On the contrary, actors relying on untrustworthy news sources showed a lower likelihood of employing trustworthy ones, suggesting that their influence was more targeted by using only untrustworthy sources. This pattern remained consistent across all three events studied. The utilization of multiplex networks and their structural measures has unveiled novel insights into the characterization of influential actors within social media landscapes.

\bibliographystyle{plain}
\bibliography{ref}

\begin{thebibliography}{10}

\bibitem{battistonStructuralMeasuresMultiplex2014a}
Federico Battiston, Vincenzo Nicosia, and Vito Latora.
\newblock Structural measures for multiplex networks.
\newblock {\em Physical Review E}, 89(3):032804, 2014.

\bibitem{bhowmickTemporalSequenceRetweets2019}
Ayan~Kumar Bhowmick, Martin Gueuning, Jean-Charles Delvenne, Renaud Lambiotte, and Bivas Mitra.
\newblock Temporal {{Sequence}} of {{Retweets Help}} to {{Detect Influential Nodes}} in {{Social Networks}}.
\newblock {\em IEEE Transactions on Computational Social Systems}, 6(3):441--455, 2019.

\bibitem{bovetInfluenceFakeNews2019}
Alexandre Bovet and Hernán~A. Makse.
\newblock Influence of fake news in {{Twitter}} during the 2016 {{US}} presidential election.
\newblock {\em Nature Communications}, 10(1):7, 2019.

\bibitem{chaMeasuringUserInfluence2010a}
Meeyoung Cha, Hamed Haddadi, Fabricio Benevenuto, and Krishna Gummadi.
\newblock Measuring {{User Influence}} in {{Twitter}}: {{The Million Follower Fallacy}}.
\newblock {\em Proceedings of the International AAAI Conference on Web and Social Media}, 4(1):10--17, 2010.

\bibitem{deveirmanMarketingInstagramInfluencers2017}
Marijke De~Veirman, Veroline Cauberghe, and Liselot Hudders.
\newblock Marketing through {{Instagram}} influencers: The impact of number of followers and product divergence on brand attitude.
\newblock {\em International Journal of Advertising}, 36(5):798--828, 2017.

\bibitem{garibayEntropyBasedCharacterizationInfluence2022a}
Ozlem~Ozmen Garibay, Niloofar Yousefi, Kevin Aslett, Jacopo Baggio, Erik Hemberg, Chathura Jayalath, Alexander Mantzaris, Bruce Miller, Una-May O’Reilly, William Rand, Chathurani Senevirathna, and Ivan Garibay.
\newblock Entropy-{{Based Characterization}} of {{Influence Pathways}} in {{Traditional}} and {{Social Media}}.
\newblock In {\em 2022 {{IEEE}} 8th {{International Conference}} on {{Collaboration}} and {{Internet Computing}} ({{CIC}})}, pages 38--44, 2022.

\bibitem{haluMultiplexNetworkHuman2019a}
Arda Halu, Manlio De~Domenico, Alex Arenas, and Amitabh Sharma.
\newblock The multiplex network of human diseases.
\newblock {\em npj Systems Biology and Applications}, 5(1):1--12, 2019.

\bibitem{heIdentifyingPeerInfluence2013}
Saike He, Xiaolong Zheng, Daniel Zeng, Kainan Cui, Zhu Zhang, and Chuan Luo.
\newblock Identifying {{Peer Influence}} in {{Online Social Networks Using Transfer Entropy}}.
\newblock In G.~Alan Wang, Xiaolong Zheng, Michael Chau, and Hsinchun Chen, editors, {\em Intelligence and {{Security Informatics}}}, pages 47--61. Springer, 2013.

\bibitem{hristovaKeepYourFriends2014a}
Desislava Hristova, Mirco Musolesi, and Cecilia Mascolo.
\newblock Keep {{Your Friends Close}} and {{Your Facebook Friends Closer}}: {{A Multiplex Network Approach}} to the {{Analysis}} of {{Offline}} and {{Online Social Ties}}.
\newblock {\em Proceedings of the International AAAI Conference on Web and Social Media}, 8(1):206--215, 2014.

\bibitem{huynhMeasuresDetectInfluencer2019}
Tai Huynh, Ivan Zelinka, Xuan~Hau Pham, and Hien~D. Nguyen.
\newblock Some measures to detect the influencer on social network based on {{Information Propagation}}.
\newblock In {\em Proceedings of the 9th {{International Conference}} on {{Web Intelligence}}, {{Mining}} and {{Semantics}}}, {{WIMS2019}}, pages 1--6. Association for Computing Machinery, 2019.

\bibitem{jainDiscoverOpinionLeader2019}
Lokesh Jain and Rahul Katarya.
\newblock Discover opinion leader in online social network using firefly algorithm.
\newblock {\em Expert Systems with Applications}, 122:1--15, 2019.

\bibitem{jainOpinionLeaderDetection2020}
Lokesh Jain, Rahul Katarya, and Shelly Sachdeva.
\newblock Opinion leader detection using whale optimization algorithm in online social network.
\newblock {\em Expert Systems with Applications}, 142:113016, 2020.

\bibitem{kitsakIdentificationInfluentialSpreaders2010}
Maksim Kitsak, Lazaros~K. Gallos, Shlomo Havlin, Fredrik Liljeros, Lev Muchnik, H.~Eugene Stanley, and Hernán~A. Makse.
\newblock Identification of influential spreaders in complex networks.
\newblock {\em Nature Physics}, 6(11):888--893, 2010.

\bibitem{liHowMultipleSocial2015a}
Weihua Li, Shaoting Tang, Wenyi Fang, Quantong Guo, Xiao Zhang, and Zhiming Zheng.
\newblock How multiple social networks affect user awareness: {{The}} information diffusion process in multiplex networks.
\newblock {\em Physical Review E}, 92(4):042810, 2015.

\bibitem{madrid-moralesWhoSetNarrative2021}
Dani Madrid-Morales.
\newblock Who set the narrative? {{Assessing}} the influence of {{Chinese}} global media on news coverage of {{COVID-19}} in 30 {{African}} countries.
\newblock {\em Global Media and China}, 6(2):129--151, 2021.

\bibitem{pengInfluenceAnalysisSocial2018a}
Sancheng Peng, Yongmei Zhou, Lihong Cao, Shui Yu, Jianwei Niu, and Weijia Jia.
\newblock Influence analysis in social networks: {{A}} survey.
\newblock {\em Journal of Network and Computer Applications}, 106:17--32, 2018.

\bibitem{rodriguez-vidalAutomaticDetectionInfluencers2019a}
Javier Rodríguez-Vidal, Julio Gonzalo, Laura Plaza, and Henry~Anaya Sánchez.
\newblock Automatic detection of influencers in social networks: {{Authority}} versus domain signals.
\newblock {\em Journal of the Association for Information Science and Technology}, 70(7):675--684, 2019.

\bibitem{saxenaEntropyBasedFlow2020a}
Chandni Saxena, M.~N. Doja, and Tanvir Ahmad.
\newblock Entropy based flow transfer for influence dissemination in networks.
\newblock {\em Physica A: Statistical Mechanics and its Applications}, 555:124630, 2020.

\bibitem{schreiberMeasuringInformationTransfer2000}
Thomas Schreiber.
\newblock Measuring {{Information Transfer}}.
\newblock {\em Physical Review Letters}, 85(2):461--464, 2000.

\bibitem{senevirathnaInfluenceCascadesEntropyBased2021}
Chathurani Senevirathna, Chathika Gunaratne, William Rand, Chathura Jayalath, and Ivan Garibay.
\newblock Influence {{Cascades}}: {{Entropy-Based Characterization}} of {{Behavioral Influence Patterns}} in {{Social Media}}.
\newblock {\em Entropy}, 23(2):160, 2021.

\bibitem{seyfosadatSystematicLiteratureReview2023b}
Seyed~Farid Seyfosadat and Reza Ravanmehr.
\newblock Systematic literature review on identifying influencers in social networks.
\newblock {\em Artificial Intelligence Review}, 56(1):567--660, 2023.

\bibitem{singhIdentificationInfluencePropagation2019a}
Niharika Singh, Aakash Malik, Oshin Maini, and Gaurav Rajput.
\newblock Identification of {{Influence Propagation Metrics}} in {{Social Networks}}.
\newblock In {\em 2019 {{International Conference}} on {{Automation}}, {{Computational}} and {{Technology Management}} ({{ICACTM}})}, pages 224--227, 2019.

\bibitem{versteegInformationtheoreticMeasuresInfluence2013}
Greg Ver~Steeg and Aram Galstyan.
\newblock Information-theoretic measures of influence based on content dynamics.
\newblock In {\em Proceedings of the Sixth {{ACM}} International Conference on {{Web}} Search and Data Mining}, pages 3--12. ACM, 2013.

\bibitem{wangMultiPlatformAnalysisPolitical2021}
Yuping Wang, Savvas Zannettou, Jeremy Blackburn, Barry Bradlyn, Emiliano De~Cristofaro, and Gianluca Stringhini.
\newblock A {{Multi-Platform Analysis}} of {{Political News Discussion}} and {{Sharing}} on {{Web Communities}}.
\newblock In {\em 2021 {{IEEE International Conference}} on {{Big Data}} ({{Big Data}})}, pages 1481--1492, 2021.

\bibitem{zareiCharacterisingDetectingSponsored2020a}
Koosha Zarei, Damilola Ibosiola, Reza Farahbakhsh, Zafar Gilani, Kiran Garimella, Noël Crespi, and Gareth Tyson.
\newblock Characterising and {{Detecting Sponsored Influencer Posts}} on {{Instagram}}.
\newblock In {\em 2020 {{IEEE}}/{{ACM International Conference}} on {{Advances}} in {{Social Networks Analysis}} and {{Mining}} ({{ASONAM}})}, pages 327--331, 2020.

\bibitem{zhuResearchKnowledgeDissemination2022a}
Hongmiao Zhu, Yumie Wang, Xin Yan, and Zhen Jin.
\newblock Research on knowledge dissemination model in the multiplex network with enterprise social media and offline transmission routes.
\newblock {\em Physica A: Statistical Mechanics and its Applications}, 587:126468, 2022.

\end{thebibliography}

\end{document}